\documentstyle[emulateapj]{article}

\font\smcap=cmcsc10

\def\simlt{$\lesssim$}
\def\simgt{$\gtrsim$}
\def\hi{{\smcap H$\,$i}}

\def\He{{\smcap H}e}
\def\Ha{H$\alpha$}
\def\HaN{H$\alpha$ + [{\smcap N$\,$ii}]}
\def\NHI{$N_{HI}$}
\def\MHI{$M_{HI}$}

\def\MhLb{$M_{HI}/L_B$}
\def\MhLr{$M_{HI}/L_R$}
\def\ML{$M_\odot \, L_\odot^{-1}$}

\def\Myr{$M_\odot \, {\rm yr}^{-1}$}
\def\mB{$\mu_B$}
\def\mR{$\mu_R$}
\def\msqas{mag arcsec$^{-2}$}
\def\col#1{$\times 10^{#1} {\rm cm}^{-2}$}
\def\B{$B$}

\def\R{$R$}

\def\about{$\sim$}

\def\Mo#1{$\times 10^{#1} M_\odot$}

\def\as{\arcsec}
\def\kms{km s$^{-1}$}
\def\ie{{\it i.e.}}
\def\eg{{\it e.g.}}
\def\et{{\it et~al.}}
\def\cf{{\it cf.}}
\def\x{$\times$}

\def\FWHM{$\theta_{FWHM}$}

\lefthead{Hibbard \& Yun}
\righthead{Tidal Tail in Arp 299}
\submitted{Accepted for publication in The Astronomical Journal}
\begin{document}

\title{A 180 Kpc Tidal Tail in the Luminous Infrared 
Merger Arp 299}

\author{J.E.~Hibbard} 
\affil{National Radio Astronomy Observatory\footnote{The National Radio
Astronomy Observatory is a facility of the National Science Foundation
operated under cooperative agreement by Associated Universities, Inc.},
520 Edgemont Road, Charlottesville, VA, 22903; jhibbard@nrao.edu}

\author{M.S.~Yun}
\affil{National Radio Astronomy Observatory$^1$, P.O.\ Box 0, Socorro, 
New Mexico, 87801; myun@nrao.edu}

\begin{abstract}

We present VLA {\hi} observations and UH88\as\ deep optical \B- and
\R-band observations of the IR luminous merger Arp 299 (= NGC 3690 + IC
694).  These data reveal a gas-rich ($M_{HI}=3.3\times 10^9 M_\odot$)
optically faint (\mB\simgt 27 \msqas, \mR\simgt 26 \msqas) tidal tail
with a length of over 180 kpc.  The size of this tidal feature
necessitates an old interaction age for the merger (\simgt 750 Myr
since first periapse), which is currently experiencing a very young
star burst (\simlt 20 Myr).  The observations reveal a most remarkable
structure within the tidal tail: it appears to be composed of two
parallel filaments separated by \about 20 kpc.  One of the filaments
is gas rich with little if any starlight, while the other is gas
poor. We believe that this bifurcation results from a warped disk in
one of the progenitors.  The quantities and kinematics of the tidal
{\hi} suggest that Arp 299 results from the collision of a retrograde
Sab-Sb galaxy (IC 694) and a prograde Sbc-Sc galaxy (NGC 3690) that
occurred 750 Myr ago and which will merge into a single object in
$\sim$ 60 Myr. We suggest that the present IR luminous phase in this
system is due in part to the retrograde spin of IC 694.  Finally, we
discuss the apparent lack of tidal dwarf galaxies within the tail.

\end{abstract}


\section{Introduction}
\label{sec:intro}

Arp 299 is a nearby ($V_{hel}$=3080 \kms)\footnote{All radial
velocities quoted in this paper are heliocentric.} peculiar system
comprised of two highly distorted disk galaxies, with NGC 3690 (= Mrk
171A = UGC 6471 = VV118b) to the west and IC 694 (= Mrk 171B = UGC
6472 = VV118a)\footnote{There is some confusion as to the naming of
this system, which is addressed separately in the appendix.  In the
following, we following the naming convention used most widely in the
literature.} to the east.  There is also a compact spheroidal galaxy
lying approximately 1$^\prime$ to the NW which is a the same velocity as
Arp 299 (MCG+10-17-2a, $V_{hel}$=3100 \kms, Fairall 1971, Sargent
1972).  Figure~\ref{fig:a299B} presents a \B-band image of this
system, taken from the optical data described later in this paper.
Each of the above named systems are labeled, as is the unrelated
background bridge-tail-bar system Arp 296.
 
Arp~299 appears to be in an advanced stage of merging, with the two
disks in contact, but the nuclei still separated by about 20\as\ (4.7
kpc)\footnote{Adopting the distance of 48 Mpc from Sanders, Scoville
\& Soifer (1991), which assumes $H_o$=75 km s$^{-1}$ Mpc$^{-1}$ and
the Virgocentric flow model of Aaronson \et\ (1982).  At this distance
1$^\prime$=14 kpc.}.  This impression is supported by radio continuum and
near-infrared (NIR) imaging (Gehrz, Sramek \& Weedman 1983) and
millimeter spectral line observations (Casoli \et\ 1989, 1999; 
Sargent \& Scoville 1991, Aalto \et\ 1997), which locate two major
concentrations of molecular gas at peaks in the radio and NIR (regions
{\bf A} and {\bf B} in Fig.~\ref{fig:a299B}, using the naming
convention of Gehrz, Sramek \& Weedman 1983).  Source {\bf A} is
identified with the nucleus of IC 694, while source {\bf B} is
identified with the nucleus of NGC 3690.\footnote{High resolution NIR
imaging by Wynn-Williams \et\ (1991) shows that source {\bf B} is
itself a double, separated by about 3\as.} Another concentration of
molecular gas occurs at the region of disk overlap (labeled {\bf
C-C}$^\prime$ in Fig.~\ref{fig:a299B}).

Spectroscopic observations of the NIR peaks in radio recombination
lines (Anantharamaiah \et\ 1993; Zhao \et\ 1997), the Mid infrared
(Dudley \& Wynn-Williams 1993; Dudley 1998), the NIR (Gehrz, Sramek \&
Weedman 1983; Beck, Turner \& Ho 1986; Nakagawa \et\ 1989; Ridgeway,
Wynn-Williams \& Becklin 1994; Doherty \et\ 1995; Smith \et\ 1996;
Lan\c con, Rocca-Volmerang \& Thuan 1996), optical recombination lines
(Bushouse \& Gallagher 1984; Keel \et\ 1985; Armus, Heckman \& Miley
1989) and in the UV (Augarde \& Lequeux 1985; Kinney \et\ 1993;
Roberts 1996; Vacca \et\ in preparation) fail to reveal any definitive
evidence for an AGN, and are consistent with the dominant energy
source being a compact starburst (although there is evidence for some
contribution due to a compact source; see Shier, Rieke \& Rieke 1996;
Carral, Turner \& Ho 1990; Jones \et\ 1990; Jones 1997).  This picture
is consistent with continuum observations in the radio (Condon \et\
1982; Gehrz, Sramek \& Weedman 1983; Condon \et\ 1990, 1991; Smith,
Lonsdale \& Lonsdale 1998), far infrared (Joy \et\ 1989), NIR
(Telesco, Decher \& Gatley 1985; Carico \et\ 1990; Wynn-Willams \et\
1991; Miles \et\ 1996), and X-ray (Rieke 1988, Zezas \et\ 1998).

Within the main body, vigorous star formation is taking place, with an
inferred star formation rate (SFR) of \about 50 \Myr\ (Heckman, Armus
\& Miley 1990), and an age of \simlt 20 Myr (Augarde \& Lequeux 1985; 
Stanford 1989; Nakagawa \et\ 1989; Meurer \et\ 1995; Vacca \et\ in
preparation).  A deep \HaN\ image of this system reveals a
``magnificent large-scale filamentary structure made up of arcs and
loops that surround the galaxies at radii $\sim$ 6--12 kpc" (Armus,
Heckman \& Miley 1990).  Recent x-ray observations reported by Heckman
\et\ (1998) show evidence for hot gas emerging from the inner regions
to the north, which the authors interpret as evidence for a hot,
expanding superwind fluid.

The present Very Large Array (VLA) 21cm observations of the neutral
hydrogen ({\hi}) in Arp 299 were conducted as part of a larger program
to study the closest luminous IR galaxies in the IRAS Bright Galaxy
Sample of Sanders \et\ (1991).  Arp 299 is both the most IR luminous
member of this sample within a distance of 50 Mpc
($L_{IR}=8.1\times10^{11} L_\odot$), and the closest system with an IR
luminosity greater than 5$\times 10^{11} L_\odot$.  It is therefore an
important target to study for insight into the causes of such luminous
starbursts.  The purpose of the present study is to map any extended
atomic hydrogen in order to constrain the encounter geometry and the
Hubble types of the progenitors.

The paper is organized in the following manner.  In \S \ref{sec:obs}
we briefly describe the observations.  In \S \ref{sec:results} we
present the observations separately for the inner regions, tidal
regions, and for the {\hi} absorption and a new {\hi} detected
companion.  In \S \ref{sec:disc} we use the observations to explore
the probable Hubble types of the progenitors, the encounter geometry,
the possibility of self-gravitating entities within the tail, the
conditions for an IR luminous phase, and the explanation for the
differing {\hi} and optical tidal morphologies.  We summarize our main
conclusions in \S \ref{sec:concl}.

\section{Observations}
\label{sec:obs}

\subsection{HI Observations}

The VLA {\hi} spectral line observations were obtained as part of a
study on the outer gas dynamics of IR luminous mergers (Hibbard \& Yun
1996 and in preparation), 
and the data reduction techniques are fully described in
that paper, except for the continuum subtraction which is described
below.  The details of these observations are tabulated in
Table~\ref{tab:HIobs}.  Briefly, the data consist of a 3.5 hour
observation with the VLA in the C-array configuration (full-width at
half maximum resolution \FWHM \about 15\as) and a 3 hour observation
with the VLA in the D-array configuration (\FWHM \about 45\as).  The
data were combined in the UV plane to form the C+D array data that
are used in the remainder of this paper. The correlator mode was
chosen to provide 63 spectral channels with a channel spacing of 10.5
\kms\ and total bandwidth of 660 \kms\ centered at 3080 \kms.  The
data were reduced using standard reduction procedures in the
Astronomical Image Processing System (AIPS).

The observations reveal broad {\hi} absorption against the disk-wide
starburst and central radio continuum sources (see also Dickey 1982;
Baan \& Haschick 1990 hereafter BH90).  This absorption was observed
in channels 5--55 out of the 63 channel data cube (of which only
channels 2--58 are useful).  As a result, we could not achieve an
optimal continuum subtraction by using line free channels on either
end of the bandpass, as is usually done (e.g.~Rupen 1998).  
Instead, we mapped and ``cleaned" the continuum image and
subtracted the brightest 17 ``clean components" from the UV line data,
which accounts for approximately half of the total continuum flux of
820 mJy falling within the primary beam (FWHM=30$^\prime$).  The line data
were then mapped and cleaned in AIPS, resulting in a three dimensional
{\hi} data ``cube", with Right Ascension and Declination along the
first two coordinates and velocity along the third axis.

These data still contained many continuum sources, but were relatively
free of the deep sidelobes from the brighter sources.  The residual
continuum was removed in the map plane using a spatially variable
continuum baseline.  The {\hi} tidal features appeared over a narrow
range of velocity, and the residual continuum was removed from these
regions by fitting a first order polynomial to a large number of
channels on either side of this range.  The emission from the NW disk
appears in channels 4--26, and the residual continuum was subtracted
from this region by fitting to the continuum in channels 34--54.  The
emission from the SW disk appears in channels 24--50, and the residual
continuum was subtracted from this region by fitting to the continuum
in channels 4--20.  Finally, the residual continuum was removed from
the regions showing {\hi} absorption by fitting to the continuum in
channels 2--4 and 52--58.

The data were mapped using different weighting functions to achieve
varying resolutions and sensitivities.  A high resolution data cube was
made using a ``Robust'' parameter ($R$; Briggs 1995) of $R$=0,
which gives more weight to longer baselines and hence to smaller angular
scales.  This cube obtains a surface brightness sensitivity of 0.35 mJy
beam$^{-1}$ at a resolution of \FWHM=17\as\x 15\as.  This corresponds to
a column density limit (2.5$\sigma$) of 4\col{19} averaged over the beam
width of 4.0\x3.4 kpc$^2$.  A more sensitive intermediate resolution
data cube was made with $R=1$, giving a resolution of 22\as\x 20\as\
(5.1\x4.7 kpc$^2$) and a column density limit of 2\col{19}.  To further
increase sensitivity to extended low column density gas, a low
resolution data cube was made by convolving the $R$=1 data cube to a
resolution of 35\as\ (8.1 kpc), reaching a detection limit of
1\col{19}. These data will be referred to in the following as the
high, intermediate, and low resolution data, respectively.

\subsection{Optical Observations}

The VLA observations reveal a gaseous stream of material extending to
a radius of 9$^\prime$ that had no known optical counterpart on existing
images.  As such, this region was targeted for deeper optical
observations with the University of Hawai'i 88\as\ telescope at Mauna
Kea Observatory.  The observational parameters are listed in
Table~\ref{tab:OPTobs}.  The Tek 2048 CCD was used with the f/10
re-imaging optics, giving a plate scale of 0.22\as\ pixel$^{-1}$ and a
field of view of 7\farcm5.  Three overlapping 600 sec \R-band images of
the region of interest were obtained in January 1995.  In January 1997
we obtained a 600 sec \B-band image centered on the main body and two
overlapping 900 sec \B-band images of the {\hi} tail.  The total
imaged region covers an area of 8\farcm7\x 12\farcm4.  The conditions
were photometric on both dates, and the seeing was \about 1\as.  The
data were calibrated via observations of Landolt UBVRI standards
(Landolt 1983) observed on the same nights, with zero point errors
(1$\sigma$) of 0.01$^m$ in \B\ and 0.03$^m$ in \R.

The images were flattened and combined using the techniques described
in Hibbard \& van Gorkom (1996, hereafter HvG96), and the final
mosaics are flat to better than one part in 500.  The combined image
was transformed to the World Coordinate System (and thereby to the same
reference frame as the radio images) by registering to an image of the
same area extracted from Version II of the Digitized Sky
Survey\footnote{The Second Palomar Observatory Sky Survey was made by
the California Institute of Technology with funds from the National
Science Foundation, the National Geographic Society, the Sloan
Foundation, the Samuel Oschin Foundation, and the Eastman Kodak
Corporation.  The Oschin Schmidt Telescope is operated by the
California Institute of Technology and Palomar Observatory.}~(XDSS),
obtained from the CD jukebox at the Canadian Astronomy Data
Centre.\footnote{The Canadian Astronomy Data Center is operated by the
Dominion Astrophysical Observatory for the National Research Council
of Canada's Herzberg Institute of Astrophysics.}~~The registration was
accomplished by referencing the location of \about 20 stars in common
on both images using the ``koords" program in the Karma software
package (Gooch 1995) to perform a non-linear least squares fit for the
transformation equations.  The plate solution so found is better than
a fraction of a pixel (0.22\as), and the overall registration should
be as good as that of the northern portion of the Guide Star Catalog,
which is estimated to be $0\farcs5$ (Taff \et\ 1990).

Finally, a deep optical image was constructed by binning the pixels with
the lowest light levels 9x9 to achieve a limiting surface brightness of
\mB=28.2 \msqas\ and \mR=27.1 \msqas\ (2.5$\sigma$; Table~\ref{tab:OPTobs}). 

\section{Results}
\label{sec:results}

The luminosity for the main galaxies was
calculated out to the \mB=26 \msqas\ isophote, and the corresponding
\B\ magnitude for the Arp 299 system ($m_B$=12.31) agrees within
0.01$^m$ to that given by Mazzarella \& Boroson (1993). These and other
global parameters for the Arp 299 system are given in 
Table~\ref{tab:global}, and the entries are described in the notes to
that table.  In the following, we will discuss the kinematics and
morphology of the inner disk first, and then that of the outer tidal
features.  We also discuss the central {\hi} absorption and report the
discovery of a small dwarf companion to Arp 299.

\subsection{Inner Gas Disk \label{sec:innerdisk}}

The disk {\hi} kinematics are illustrated in Figures~\ref{fig:innerLV}
\& ~\ref{fig:innerChan}.  Fig.~\ref{fig:innerLV} presents several
2-dimensional representations of the high resolution {\hi} data, with
the emission summed over the third dimension.  The upper left panel
shows the integrated intensity map in solid contours (lowest contour
corresponds to \NHI\about 5\col{19}) upon the \B-band image.  This map
emphasizes higher column density material at the expense of the
diffuse lower column density {\hi}.  We have therefore provided a
dotted contour to indicate the \NHI=2\col{19} contour of the
intermediate resolution data, in order to give a more complete picture
of the {\hi} distribution in the inner regions.  We have also labeled
a minor axis clump of {\hi} and a western plume for reference in the
following discussion.

To the right and below this panel are position--velocity maps
constructed by summing the emission over declination (lower panel) or
right ascension (right panel).  Fig.~\ref{fig:innerChan} presents the
high resolution channel maps of the inner disk alone, after Hanning
smoothing by a factor of two in velocity.  In both of these maps,
absorption is indicated by white dashed contours.

The inner {\hi} disk was previously mapped with the VLA in the 
C-array configuration by Stanford \& Wood (1989).  As discussed by these
authors, the kinematic signature seen in Figs.~\ref{fig:innerLV}
\& \ref{fig:innerChan} is representative of a rotating disk, with
the NW half of the disk receding and the SW half approaching.  The NW
component of the disk reaches a radius of 25 kpc and contains
2.9\Mo{9} of {\hi}.  The SE component of the {\hi} disk reaches a
radius of 21 kpc and contains a total of 2.8\Mo{9} of atomic
hydrogen.\footnote{These masses do not include the contributions due
to gas in the western plume or minor axis clump discussed later.} The
disk spans the velocity range from 2880--3390
\kms, with its kinematic center at 3135 \kms.

This disk was also mapped by the VLA in the D-configuration by Nordgren
\et\ (1997) using a coarser velocity resolution (by a factor of four)
and a larger bandpass (by a factor of two) than the observations
presented here.  Their observations suggest that the emission from the
NW disk continues to 3550 \kms\ (see their Fig.~\ref{fig:TailOpt}b),
and therefore it is likely that we have missed some disk emission due
to our smaller velocity coverage.  The OH maser emission from Arp 299
(Baan 1985; BH90) suggests that there may be nuclear gas at velocities
as high as 3650 \kms.

The {\hi} disk appears to be physically associated with the optically
distorted disk of IC 694.  This view is strongly supported by both the
CO and \Ha\ kinematics mapped at higher spatial resolution, as will be
discussed in \S \ref{sec:encounter} below.  There are, however,
emission features which deviate from this simple disk motion.  There
are 4.6\Mo{8} of gas in a western extension to the NW portion of the
disk, and which appears kinematically associated with that disk (\ie,
appears in the same velocity channels in Fig.~\ref{fig:innerChan}),
but which is spatially associated with a faint plume of starlight
reaching to the SW (labeled ``western plume" in
Fig.~\ref{fig:innerLV}, see also Fig.~\ref{fig:a299B}).  There are also
1.5\Mo{8} of gas in a clump lying along the SW minor axis of the disk
(labeled ``minor axis clump" in Fig.~\ref{fig:innerLV}).  
If this minor-axis material were associated with the disk of IC 694, 
it should appear in only a few spectral channels.
Since it appears over a broad range of velocities (from
3050---3200 \kms\ in Fig.~\ref{fig:innerChan}), we suggest instead
that it is associated with NGC 3690 (see \S \ref{sec:encounter}).
There is also some emission associated with the base of the tidal tail
(seen in the right panel of Fig.~\ref{fig:innerLV} towards the north
and in the channel maps from 3092---3135\kms), which is seen more
clearly in the low resolution maps discussed in the next section.
Finally, there are some {\hi} features that project close to the compact
spheroidal galaxy MCG+10-17-2a (Fig.~\ref{fig:innerChan}, channels at
3306---3371 \kms).  Since these features appear at velocities that are
higher than the velocity of MCG+10-17-2a itself ($V_{hel}$=3100
\kms, Fairall 1971, Sargent 1972), and close to the velocites of 
the {\hi} within the underlying disk of IC 694, it seems likely that
they are due to disk material that was perturbed by a recent close
passage of the spheroidal.

\subsection{Outer Tidal Morphology \label{sec:outerMorph}}

Our VLA observations revealed a previously unknown tidal tail reaching
124 kpc in radius to the north.  This discovery was initially reported
in Hibbard \& Yun (1996).  Subsequently, VLA D-array observations of
Arp 299 from May 1995 were published by Nordgren \et\ (1997),
confirming both the {\hi} and optical tidal tail we report on in this
paper.

In Figure~\ref{fig:outerLV} we show the {\hi} integrated intensity map
from the intermediate resolution data, with position-velocity profiles
plotted along either axis as in Fig.~\ref{fig:innerLV}.  The tail
appears as two parallel filaments separated by about 20 kpc (labeled
``inner filament" and ``outer filament" in Fig.~\ref{fig:outerLV}),
extending northward and connecting onto or merging into a dense clump
of gas (labeled ``N clump" in Fig.~\ref{fig:outerLV}).  The inner
filament originates near the minor axis of IC 694 (see also the
northernmost clump in Fig.~\ref{fig:innerLV}), appears broken at the
lowest contour drawn in Fig.~\ref{fig:outerLV} (\NHI\about
2.5\col{19}) after which there appears an {\hi} density peak
($N_{HI,peak}\sim$ 2\col{20}, labeled ``{\hi} knot" in
Fig.~\ref{fig:outerLV}), and the filament continues northward.  This
inner filament contains a total of 4.4\Mo{8} of neutral atomic
hydrogen.  The outer filament originates to the west of the first (see
also the second northernmost clump in Fig.~\ref{fig:innerLV}), is more
complete and of a higher average column density than the inner
filament (\NHI\about 1\col{20}), and contains twice as much {\hi}
(\MHI=9\Mo{8}).  The filaments join at the N clump, which contains
2\Mo{9} of {\hi}.  This clump has an irregular column density
distribution with peaks as high as 3\col{20}.  The kinematics of these
features will be discussed in the next section.

The tail reaches a maximum projected radial separation of 8\farcm6
(124 kpc; measured from source {\bf A} to the outer contour in
Fig.~\ref{fig:outerLV}), and extends 12\farcm7 or 180 kpc as measured
along its length (measured from source {\bf A} along the peak {\hi}
column density contour of the outermost feature).  It contains a total
of 3.3\Mo{9} of atomic hydrogen, which is at the upper end of the
tidal gas content of other major disk-disk mergers (HvG96, Hibbard \& 
Yun in preparation).  Since less than half of the progenitor disk material can 
be raised into a tail, and most of this falls back quickly onto the
remnant (Hibbard \& Mihos 1995), this amount of tidal
{\hi} requires a very {\hi} rich progenitor ($M_{HI} \sim 10^{10}
M_{\odot}$).  This point will be explored in more detail in \S
\ref{sec:progen}.

The deep optical imaging reveals a faint stellar tail extending in the
same direction as the {\hi} features (see Figures \ref{fig:DeepOpt},
\ref{fig:a299rgb} \& \ref{fig:TailOpt}).  The optical tail appears to 
emerge in the north near the minor axis of the system and curve
slightly to the NE to a radius of 9$^\prime$.  The width of the
optical tail appears remarkably constant at \about 1$^\prime$ (14
kpc), although there may be some widening in the \B-band image.  The
observed width of the tail can be understood as the result of the
dispersal of outer disk stars with a typical velocity dispersion of 20
\kms (e.g.~van der Kruit 1988; Bottema 1993) over an interaction age
of 750 Myr (\S \ref{sec:encounter}).  At very faint light levels the
tail appears to loop south and west from the NE tip towards the bright
star labelled ``Star 2'' in Fig.~\ref{fig:DeepOpt}a, with perhaps a
very faint extension directly south of this star.  While scattered
light from the stars labelled 1 \& 2 in this figure make the precise
morphology of the faint optical light uncertain, these features have
the same morphology in both the \B- and \R-band greyscale images of
Fig.~\ref{fig:DeepOpt}, while the scattered light properties are
different between the bands (see e.g.~the difference between the faint
scattered light associated with Star 3 in Figs.~\ref{fig:DeepOpt}a \&
\ref{fig:DeepOpt}b), and we feel confident that the faint optical
features are not artifacts.

Upon first examination the optical and {\hi} tidal features appear
similar.  But when the data are overlaid, as in Figs.~\ref{fig:a299rgb} 
\& \ref{fig:TailOpt}b, the differences in the distribution are striking: 
the inner {\hi} filament roughly aligns with the optical tail, while the 
outer parallel {\hi} filament has no corresponding optical feature down 
to the faintest levels measured (\mB \about 28.5 \msqas, \mR \about 27.5 
\msqas).

Even within the optical tail, there is a poor correlation between
optical and gaseous density peaks.  If anything, the gas peaks seem to
be displaced a bit to the west of the optical peaks.  Neither the gap
nor the knot in the inner {\hi} filament appear to coincide with any
discrete optical feature, although the optical surface brightness
drops by \about 1 \msqas\ just north of the knot
(Fig.~\ref{fig:DeepOpt}).  The displacements are even more extreme in
the northernmost regions of the tail, where the gas and starlight
appear to be anti-correlated: to either side of the optical tail there
are regions of high gas column density (\NHI\about 3\col{20} to the
west, \NHI\about 2\col{20} to the east), whereas the gas within the
tail has \NHI\about 1\col{20}. At its northernmost extent, the optical
tail has no associated {\hi}, and actually appears to ``hook" towards
the east, {\it exactly} around the northern {\hi} contours in
Fig.~\ref{fig:TailOpt}b.

This anti-correlation suggests the possibility that dust associated
with the {\hi} gas is playing a role in shaping the optical tidal
morphology.  However the required amount of extinction ($>$1 mag at
\NHI = 2\col{20}) is an order of magnitude larger than measured for
Galactic {\hi} (Bohlin \et\ 1978).  Since tails are drawn from the
outer, presumably dust-poor regions of disks, such a high gas-to-dust
ratio seems unlikely (see also Alton \et\ 1998).  Evidence for just
such a high gas-to-dust content in tidal gas may be indicated by
observations of red globular clusters behind an HI tidal stream in NGC
5018 (Hilker \& Kissler-Patig, 1996).  Still, we feel that this
explanation is ruled out for Arp 299 by the fact that the optical tail
is actually bluer north of the {\hi} knot, where the {\hi} column
density is highest, compared to the starlight in the ``gap" in the
inner {\hi} filament, where the {\hi} column density is lowest
(Fig.~\ref{fig:TailOpt}a).  One would expect the opposite trend if
there was a high gas-to-dust ratio in this tidal gas (however see
Witt, Thronson \& Capuano 1992).  The $B-R$ colors of the stellar tail
range from 0.3--1.8, similar to range found in the outer stellar disk
of late type galaxies (de Jong 1995, Ch.~4) and in other tailed
mergers (Schombert \et\ 1990; Hibbard \et\ 1994; Hibbard 1995).

The total stellar content of the optical tail was estimated by masking
the optical images below \mB = 28 \msqas and \mR = 27 \msqas\ and
replacing regions containing point sources (presumably background
sources, \S \ref{sec:Tidwf}) and the optical halos of Stars 1\& 2 in
Fig.~\ref{fig:DeepOpt}a 
with adjacent background values.  The total optical light in the
northern tail is $L_B=1.6\times 10^9 L_{\odot,B}$ and $L_R=1.2\times
10^9 L_{\odot,R}$, corresponding to 4\% and 2\% of the total \B- and
\R-band light of the system, respectively\footnote{These numbers
increase to $L_B=2.1\times 10^9 L_{\odot,B}$ (5\%) and $L_R=1.9\times
10^9 L_{\odot,R}$ (4\%) if the background point sources are left in
and a 0.5 \msqas\ lower surface brightness threshold is used.}
(Table~\ref{tab:global}). These percentages are on the lower end of
the values found in peculiar systems (Schombert \et\ 1990; Hibbard
\et\ 1994; HvG96).

The total {\hi} mass-to-blue light ratios (in solar units) for all the
northern tidal features is \MhLb = 1.8.  There is quite a variation
within the tail, with \MhLb\ \about 0--1 along the optical tail, and
ranging from 2 to over 10 for the outer filament and eastern regions
of the N clump.
The faint plume of light extending 45 kpc in radius to
the west of Arp 299 contains $L_B=1\times 10^9 L_{\odot,B}$.  
Associating the western {\hi} extension to this feature gives \MhLb=0.5.

\subsection{Outer Tidal Kinematics \label{sec:outerKin}}

The kinematics of the tidal regions are illustrated by the integrated
position-velocity profiles in Fig.~\ref{fig:outerLV}, intensity
weighted velocities and velocity dispersions in
Fig.~\ref{fig:TailOpt}c\&d, individual channel maps in
Fig.~\ref{fig:outerChan}, and two position-velocity slices through
either {\hi} filament in Fig.~\ref{fig:filamentLV}.  Both
Fig.~\ref{fig:outerChan} and Fig.~\ref{fig:filamentLV} are constructed
from the low resolution data cube, and the lowest plotted contour
corresponds to a column density of 1\col{19}.  The channel maps are
only plotted over the velocity range containing tail emission.

The kinematics immediately illustrate two important points.  The first
is that the tidal kinematics are smooth, continuous and single valued,
as expected for tidal features (e.g.  Barnes 1988; Hibbard \et\ 1994;
Hibbard \& Mihos 1995).  The second is that the two {\hi} filaments s
hare the same
kinematics.  This is best seen in Fig.~\ref{fig:outerChan}, in which
the {\hi} emission associated with parallel regions along the
filaments appears in identical channels all along the tail.  This
rules out the possibility that the two filaments are separate tidal
tails that are simply projected near each other; they must be
different regions of the same kinematic structure.  The figures also
show that the N clump is continuous with the filaments and not a
distinct entity.  Therefore, the N clump and two filaments form a
single contiguous feature in both space and velocity.

A more detailed comparison of the filament kinematics is shown in
Fig.~\ref{fig:filamentLV}, where we have plotted position-velocity
profiles along both.  This figure shows that the base of the outer
filament has a slightly higher mean velocity than the base of the
inner filament, which may help in deciding the relative morphologies,
although we await detailed kinematic modeling before deciding how to
interpret this information.  Fig.~\ref{fig:filamentLV} illustrates a
few more interesting properties of the filaments.  In
Fig.~\ref{fig:filamentLV}b we see that the apparent ``gap" in the
integrated intensity map of Fig.~\ref{fig:outerLV} is not 
devoid of {\hi}: weak emission which did not pass the threshold used 
when making the moment maps bridges this gap (see also the panel at 
3150 \kms\ in Fig.~\ref{fig:outerChan}).  We
also see that the ``knot" within the inner filament has an anomalously
large velocity width (see also line-width map in
Fig.~\ref{fig:TailOpt}d): the majority of the tail has a velocity
dispersion of 7--9 \kms, which is typical of tidal tails in general
(Hibbard \et\ 1994, HvG96), while the knot has a dispersion of over
twice this (20 \kms). Fig.~\ref{fig:TailOpt}d shows that there are two
additional velocity dispersion peaks along the southern edge of the N
clump, with dispersions of 17 \kms\ and 13.5 \kms\ compared to an
average dispersion of 10.5 \kms\ within the N clump. None of these
regions corresponds to any noticeable features in the optical images
or color maps, which argues against their being related to the
putative tidal dwarf galaxies, a question we investigate in more
detail in \S \ref{sec:Tidwf}.

Finally, there is a high velocity envelope of material just north of
the knot (indicated in Fig.~\ref{fig:filamentLV}b).  A similar envelope
of material is seen at parallel locations in the outer filament
(Fig.~\ref{fig:filamentLV}d).  The channel maps show that this gas
lies exactly along the faint optical light that stretches from the end
of the tail towards Star 2 in Fig.~\ref{fig:DeepOpt}a (see panels at
3171--3161 \kms\ in Fig.~\ref{fig:outerChan}), while the majority of
the filament gas has lower velocities and lies north of this light
(see panels at 3098--3140 \kms\ in Fig.~\ref{fig:outerChan}).  These
are probably important clues to the origin of these filaments,
although we are at a loss to say what.

\subsection{Central Absorption \label{sec:absorp}}

High resolution (\about 5\as) VLA radio continuum observations show
that half of the total the 1.4 GHz radio continuum emission lies
within 5\as\ of regions {\bf A, B} and {\bf C-C}$^\prime$
in Fig.~\ref{fig:a299B}, with the other half of the flux distributed
between these regions and throughout the disk of IC 694 (Gehrz, Sramek
\& Weedman 1983).  Even higher resolution (\FWHM=$1\farcs7$) VLA
A-array observations of the {\hi} absorption against these continuum
sources are presented in BH90, along with observations of OH megamaser
emission.  These observations show that {\hi} absorption is seen
against both the compact continuum sources as well as against the
extended continuum features.  Our continuum profiles are quite similar
to those published by BH90, with subtle differences that are 
likely due to our larger beam, which allows surrounding {\hi} emission to
fill in some of the absorption.  We will therefore use the results of
the higher resolution {\hi} absorption observations in concert with
our {\hi} emission observations to set some rough limits on the spin
temperature (T$_{\rm spin}$) of the {\hi} gas.  Along the way, we will
estimate the absorbing column of gas in front of the continuum
sources, from which we will derive an approximate correction to the
total {\hi} mass.

BH90 found that the {\hi} absorption is dominated by a broad component
($\Delta V \sim$120 \kms) at $V_{hel}$=3160 \kms, and a narrow
component ($\Delta V \sim$70 \kms) at $V_{hel}$=3050 \kms.  Due to the
near correspondence of these velocities with the systemic velocities
of IC 694 and NGC 3690 (3080--3180 \kms and 3000--3080 \kms,
respectively; BH90 and reference therein) and because these two
components are found in the absorption spectrum against {\it all} of
the radio continuum sources, BH90 deduce that they are due to outer
disk material from both IC 694 and NGC 3690 lying in the foreground
rather than near-nuclear gas clouds.  This allows us to interpolate
our {\hi} column density maps across the absorbing regions. This
column density can then be used together with the ratio of the neutral
gas absorbing column density to the spin temperature ($N_{HI}/T_{\rm
spin}$) given by BH90 to constrain $T_{spin}$.

From the column densities given in Fig.~2, we estimate a foreground
column density of $N_{HI} \approx$1--2\col{21} due to the disk of
IC~694. For the corresponding absorption component (i.e.~the broad
component at $V_{hel}$=3050 \kms), BH90 find $N_{HI}/T_{\rm spin}
\sim$3--8\col{19} $f_g$ K$^{-1}$, where $f_g$ is the fraction of the
radio continuum source covered by absorbing gas.  Since BH90 find this
absorption feature in the spectra of {\it each} continuum source in
Arp 299, $f_g$ must be near unity, yielding a spin temperature of
12--70 K. This is similar to the range found in our galaxy (34--74 K;
Kalberla \et\ 1985). The lower range of temperatures are similar to
the very low values of T$_{\rm spin}$ (T$_{\rm spin} <$ 50 K) found in
the Magellanic stream (Kobulnicky \& Dickey 1999) and the nearby
galaxies NGC 247 and NGC 253 (Dickey, Brinks \& Puche 1992).

Using this range of spin temperatures and the measured $N_{HI}/T_{\rm
spin}$ values from BH90 for the low velocity absorption component
($V_{hel}$=3050 \kms, $N_{HI}/T_{\rm spin}$ = 1--2\col{19} K$^{-1}$)
we deduce a column density of $10^{20} - 10^{21} {\rm cm}^{-2}$ due to
gas from NGC 3690.  Adding these two contributions\footnote{There are
additional absorption components seen in the {\hi} absorption spectra
obtained by BH90, but their association with OH emission features at
similar velocities places them very close to the nucleus of IC~694
with a small covering factor, and they will therefore contribute very
little to the total {\hi} mass.}  gives an expected column density of
1--3\col{21} in front of the absorption region.  Doubling these values
to account for emission behind the continuum sources we conservatively
estimate an average column density in the absorbing region of
2\col{21}, corresponding to a ``missing'' {\hi} mass of 2.2\Mo{9}.
This value is used to arrive at the ``absorption corrected" \MHI\
value tabulated in Table~\ref{tab:global}.  This total is very
approximate, and could easily be 2--3 times as large (see \S
\ref{sec:progen}).

\subsection{HI Detected Companion}

The {\hi} observations uncovered a previously unknown and uncataloged
companion to Arp 299 containing $8\times 10^7 M_\odot$ of {\hi}.  This
companion lies \about 10$^\prime$ west and 3$^\prime$ south of Arp 299
(projected separation of $r$=140 kpc) and has a systemic velocity of
+145 \kms\ with respect to Arp 299. This region was not covered in our
optical observations.  An amorphous optical counterpart is found at
this location on the XDSS image.  The {\hi} column density is
contoured upon the XDSS image in Figure~\ref{fig:compMom0}, and
channel maps are shown in Figure~\ref{fig:compChan}.

We attempt to derive an optical luminosity for this companion from the
POSS-E plate of Version I of the DSS, using the photometric solution
given at {\it http://www-gsss.stsci.edu/dss/photometry/poss\_e.html}.
However, the number of counts from this source (log[counts] = 4.52) is
lower than the levels over which calibration has been reliably
determined (log[counts] = 4.8, $m_{POSS-E} < 16$ mag).  Extrapolating
the faint end of the above curve to the observed number of counts
gives a somewhat uncertain estimated magnitude of $m_{POSS-E}$=17 mag.
Assuming an average value of $B-R=$ +1.0 (e.g. van Zee, Haynes, \&
Salzer 1997), the absolute $B$ magnitude of the {\hi} companion is
about $M_B=$ $-$15.4.

The resulting global properties for the companion are given in
Table~\ref{tab:global}.  The optical appearance and {\hi} properties
of this system are similar to those of dwarf irregulars (Hoffman \et\
1996), and it is likely to have been a dwarf irregular companion of
one of the progenitors.  For comparison, it is about 20 times fainter
and 30 times less gas-rich than the LMC (Tully 1998).

\section{Discussion}
\label{sec:disc}

Although the differences between the optical and gaseous tidal
morphology are bizarre, the tail kinematics and general morphology are
very similar to the gas-rich tails of more classically understood
mergers (e.g.~Schweitzer 1978; van der Hulst 1979; Hibbard \et\ 1994;
HvG96), and the two main bodies certainly appear to be separate
disk systems that have been strongly perturbed.  It therefore seems
that the tidal hypothesis --- that these features arise as a result of
two spiral galaxies falling together under their mutual gravitational
attraction and merging due to the effects of dynamical friction
(e.g.~Toomre \& Toomre 1972 hereafter TT72; Barnes \& Hernquist 1996)
--- still provides the simplest explanation for the many peculiarities
in this system.

Given this premise, we now aim to understand the gross characteristics
of the encounter: what was the general encounter geometry (\S
\ref{sec:encounter}), and what were the approximate Hubble types of
the progenitors (\S \ref{sec:progen})?  We explore possible
explanations for the differences between the gaseous and stellar tidal
features in \S \ref{sec:morph}, although much of this discussion is
deferred to a separate paper (Hibbard, Vacca \& Yun 1999, hereafter
HVY99). In \S \ref{sec:IRphase} we examine what, if anything, the
state of this system tells us about the conditions under which super
starbursts are initiated, leading to an ultraluminous infrared phase.
Finally, in \S \ref{sec:Tidwf} we discuss the lack of putative tidal
dwarf galaxies in the outer tidal features.

\subsection{The Nature of the Encounter \label{sec:encounter}}

   The tail kinematics do not themselves determine the spin geometry 
   of their progenitor. This is because the line-of-sight velocity fields of 
   tidal features are often dominated by projection effects. This is 
   particularly true of long filamentary tails, for which the space 
   velocities will be oriented primarily perpendicular to our 
   line-of-sight (leading directly to their large apparent lengths). 
   Instead, we use the morphology of the tail to deduce the relative 
   spin geometry between its progenitor disk and the orbital plane. 
   In particular, only disks with a prograde spin geometry (\ie, a 
   spin angular momentum roughly aligned with the orbital angular 
   momentum of the two merging galaxies) can raise such long, drawn 
   out features (TT72, Barnes 1988), and the progenitor of the northern 
   tail must have had just such a prograde spin orientation. 

   Further, since tidal tails remain close to the original spin plane
   of the disk from which they are extracted (TT72), and the observed
   tail in Arp 299 emerges at nearly right angles to the disk of
   IC 694, we conclude that IC 694 could not have given rise to this 
   feature. The tail must therefore have arisen from NGC 3690. Similarly, 
   the lack of any narrow drawn-out features near the plane of the disk 
   of IC 694 suggest that this system experienced the encounter in a 
   very highly inclined or retrograde sense. Such encounters provoke 
   a much milder kinematic response in a disk (e.g. TT72; White 1979; 
   Noguchi 1991), accounting for the survival of the disk morphology 
   and kinematics of IC 694 so late into the encounter.  This spin 
   geometry agrees with the kinematics of the inner regions derived 
   from \Ha\ emission line kinematics, which show that IC 694 and 
   NGC 3690 rotate in opposite directions (Augarde \& Lequeux 1985; 
   Hibbard, Bland-Hawthorn \& Tully, in preparation).  Further, the
   systemic velocities for the two nuclei ($V_{hel}\approx$ 3135 \kms\ 
   for IC 694 [\S \ref{sec:innerdisk}] and $V_{hel}\approx$ 3040 \kms\ 
   for NGC 3690 [\S \ref{sec:absorp}])  give them the proper sense of 
   motion for the inferred spin geometry --- the nuclei rotate about each
   other opposite the direction that the disk of IC 694 rotates. 

The morphology and kinematics of the ionized gas in Arp 299 has
been mapped with Fabry-Perot interferometry (Hibbard, Bland-Hawthorn
\& Tully, in preparation), showing that there is spatial and kinematic
continuity between the outer {\hi} disk all the way inward to source
{\bf A}. This implies that the entire disk structure belongs to IC 694;
\ie\ the {\hi} is not distributed in a settled disk in rotation about
both systems, as suggested by Stanford \& Wood (1989). Further, these
observations reveal that NGC 3690 and sources {\bf B \& C-C$^\prime$}
are kinematically distinct from the IC 694 disk, as expected from the
inferred prograde-retrograde spin geometry.

From the redshifted velocities along most of the outer tail, and the
fact that the tail should be predominantly in expansion along most of
its length (Hibbard \& Mihos 1995), we deduce that the tail curves
away from us and should connect to the main bodies from behind.  In
this case, the decrease in velocities seen at the very base of the
tail in Fig.~\ref{fig:filamentLV}b (3rd arrow from the bottom in both
figures; see also northern region of upper right panel in
Fig.~\ref{fig:innerLV}), indicates that the material at the base of
the tail has reached its apocenter and is falling back to smaller
radii (e.g., Hibbard \et\ 1994; Hibbard \& Mihos 1995).  Similarly,
the broad range of velocities seen for the minor axis clump of {\hi}
(Fig.~\ref{fig:innerChan}) suggests that this material is part of the
northern tail connecting back to its progenitor, NGC 3690.  If there
is more {\hi} associated with the disk of NGC 3690, it would be
difficult to disentangle from the emission due to the SW portion of
the disk of IC 694 and the central HI absorption.

Finally, the western plume may be disk material that was
perturbed during an earlier pericentric passage of the two galaxies
about each other or by the passage of the compact NW spheroidal
(MCG+10-17-2a).  Since tidal material tends to lie within the plane it
occupies at the time it is raised (TT72), the distribution of this gas
should reflect the plane of the disk at an earlier stage of the
merger, perhaps at an early pericentric approach.  The fact that this
plane is different from the present plane suggests that the angular
momentum of the disk was changed during the encounter, evidence for
which is also seen in the \Ha\ Fabry-Perot maps at smaller radii.

These observations show the power of sensitive {\hi} and deep optical
imaging of peculiar systems.  It has long been believed that Arp 299
is in a very early stage of its interaction, given the relative youth
of the on-going starburst (\simlt 20 Myr; Vacca \et\ in preparation;
Meurer \et\ 1995; Stanford 1989; Nakagawa \et\ 1989; Augarde \&
Lequeux 1985).  The apparent lack of tidal features in their {\hi}
observations led Standford \& Wood (1989) to conclude that the outer
tidal {\hi} had quickly settled into a disk configuration early in the
encounter.  The discovery of the 180 kpc tidal tail obviously changes
this conclusion.  It also requires the interaction to be quite old.
Taking a typical rotation speed of 240 \kms\ from the inner disk
kinematics and saying that this is the maximum velocity available to
the tidal debris, we calculate that the tail has taken at least 750
Myr to form.  Therefore {\it the dynamical age of the merger is
substantially larger than the age of the present starburst}.  Arp 299
is not unique in this sense (e.g.~Mihos \& Bothun 1998), and this
cautions against using the age of the starbursts to age-date an
interacting system, as is frequently done for the peculiar
star-bursting objects at high redshift.  Further, it is possible that
Arp 299 experienced several short bursts of elevated star formation
since the tail was first launched.  In this case, the accumulated
burst populations would not be co-eval, but would show a spread of
ages, and the expected age signatures in the resulting remnant will be
much less pronounced then is often presumed (\eg\ Silva \& Bothun 1998).

\subsection{The Nature of the Progenitors \label{sec:progen}}

Now that we can use the {\hi} kinematics to approximately associate
the different kinematic features with their progenitors, we attempt to
divide the gas content and $B$-band luminosities between the two
systems. We will then use these global quantities in concert with the
statistical properties of normal Hubble types in an attempt to deduce
the approximate Hubble types of the progenitors. It should be noted
that not only are there significant uncertainties inherent in
apportioning the gas and light between the two systems, but there is
also a notoriously large scatter in global properties along the Hubble
sequence, and a further uncertainty due to conversion between gas
phases and from gas into stars due to the on-going starburst.  However
the following should provide a very rough guess for the likely
progenitors.

As mentioned above, we associate the {\hi} from the tidal tail and
minor axis clump with NGC 3690 (\MHI=3.5\Mo{9}), and the inner disk
and W plume of {\hi} with IC 694 (\MHI=6.2\Mo{9}).  Simple dynamics
suggest that less than one-half of the disk can be ejected into a
tidal tail, and that most of this quickly falls back onto the merging
system (Hibbard \& Mihos 1995).  This suggests a total {\hi} content
of 2--3 $\times M_{HI,tidal} =$ (7--10)\Mo{9} for NGC 3690.  However,
this value is already more than the combination of the tidal {\hi} and
the absorption correction (5.7\Mo{9}). This discrepancy suggests that
our absorption correction was probably too conservative (\S
\ref{sec:absorp}).  Rather than trying to account for gas that we have
no firm measure of, we simply associate all of the absorption
corrected gas (2.2\Mo{9}) with NGC 3690 and remark that the {\hi}
masses for both IC 694 and NGC 3690 are likely significantly
underestimated. Using the $12CO(1-0)$ observations of Casoli \et\
(1999), we associate the ``widespread'' molecular gas and that near
source {\bf A} with IC 694, the molecular gas near source {\bf B} with
NGC 3690, and split the molecular gas from the overlap region ({\bf
C-C$^\prime$}) between the two systems.  Total gas masses are
calculated by multiplying the {\hi} mass by 1.34 to account for \He\
and adding this to the molecular gas mass.  Finally, the individual
$B$-band luminosities are estimated by summing the light over
irregular polygons drawn around each system, and are therefore also
very approximate.  The results of this division are given in
Table~\ref{tab:progen}, and we use these values in concert with the
following statistical properties of normal Hubble types:

\begin{itemize} \item{$M_{H_2}/M_{HI}$ increases towards earlier Hubble types
(Young \& Knezek 1989, Young \& Scoville 1991).  The high values in
Table~\ref{tab:progen} suggests progenitor types later than Sab and
earlier than Sc.  They also argue against low surface brightness
progenitors (de Blok 1995)}

\item{$M_{HI}/L_B$ increases towards later Hubble types (Roberts \& 
Haynes 1994; Giovanelli \& Haynes 1990). The high values in
Table~\ref{tab:progen} suggest types later than Sa (Wardle \& Knapp
1986, Bregman \et\ 1992), and later than Sb (Roberts \& Haynes 1994).}

\item{The large mass of {\hi} suggests types later than Sa (Huchtmeier 
\& Richter 1988; Bregman \et\ 1992) and earlier than Sc (Huchtmeier
\& Richter 1988; Giovanelli \& Haynes 1990).}

\item{The large values of $M_{H_2}$ and $M_{gas}/L_B$ suggest types later
than Sa (Hogg \et\ 1993).}

\item{The large values of $M_{H_2}/L_B$ favor types later than Sab 
(Lees \et\ 1991, Bregman \et\ 1992).}

\end{itemize}

These consideration suggest progenitor types of Sab-Sc for both
systems -- an earlier type for IC 694 and a later type for the NGC
3690.  This view is supported by high-resolution (1\as) NIR imaging of
the brightest sources at 2.2 $\mu$m by Wynn-Williams \et\ (1991).  In
that study, the light profiles of the brightest NIR sources within Arp
299 were compared with stellar profiles taken immediately before and
after the on-source exposures.  It was found that source {\bf A} is
clearly resolved, with a FWHM of 460 pc (correcting to our adopted
distance of 48 Mpc), while the point sources within NGC 3690 are found
to have cores that are indistinguishable from stars, implying that
they have diameters of less than 230 pc at 2.2$\mu$m.  These
considerations suggest a more extended bulge for IC 694 than for NGC
3690, lending further support to an earlier type for the former and a
later type for the latter.  We therefore suggest progenitor types of
Sab-Sb for IC 694 and Sbc-Sc for NGC 3690.

In summary, we believe that Arp 299 results from a prograde-retrograde
encounter between a gas-rich Sab-Sb (IC 694) and Sbc-Sc (NGC 3690)
that occurred 750 Myr ago. We estimate
that the system will fully merge on a timescale somewhere between a
crossing time ($d/v_{max}$) and an orbital period ($\pi d/v_{max}$) at
the present separation $d$ of 4.7 kpc, or sometime between 20--60 Myr
from now (for a rotational speed of 240 \kms).

\subsection{Explanations for Outer Tidal Morphology \label{sec:morph}}

The differences between the gaseous and stellar tidal morphologies were
very unexpected.  Displacements between optical and gaseous tails or
bridges are not unheard of (e.g.~NGC 4725/47, Wevers \et\ 1984; Arp
295, HvG96; NGC 7714/5, Smith \et\ 1997; NGC 2782, Smith 1994); nor
are gaseous tidal tails or streamers with little, if any, starlight
(e.g.~NGC 2444/5, Appleton \et\ 1987; M81, Yun \et\ 1994); nor are
stellar tails or plumes with no {\hi} (e.g.~NGC 3921, HvG96; Arp 105,
Duc \et\ 1997; NGC 2782, Smith 1994).  There are even examples of much
less extreme ``bifurcated" gaseous tails, in which a single tidal tail
appears to split near its base into parallel {\hi} filaments, both of
which have similar kinematics (e.g.~M81 Yun \et\ 1994; NGC 3921 HvG96;
NGC 2535/6 Kaufman \et\ 1997; NGC 4038 Hibbard, van der Hulst \&
Barnes, in preparation).  What is unique in Arp 299 is that all these
effects occur at once --- there is a tail which is bifurcated along
almost its entire length, with one filament gas rich and the other gas
poor, and a striking anti-correlation between the gaseous and stellar
filaments where the filaments join. In this subsection, we investigate
possible explanations for each of these characteristics in turn.

\subsubsection{Gas-Rich Outer Filament \label{sec:gasrich}}

The high \MhLb\ ratios found for the outer filament are rather extreme,
with peaks $>$10 \ML\ and a mean of \about 4 \ML.  However they are
similar to the values found in other gaseous tidal extensions (e.g.  Arp
143 Appleton \et\ 1987; NGC 2782 Smith 1994; NGC 7252 Hibbard \et\
1994).  In these systems, the gas-rich extensions appear to be due to
the fact that \MhLb\ is an increasing function of radius in spiral
galaxies (Wevers \et\ 1986)\footnote{This is usually expressed as the
well-known fact that {\hi} disks often extend further than optical disks
e.g.  Bosma 1981, Wevers \et\ 1986, Cayatte \et\ 1994, Broeils \& van
Woerden 1994.}.  Since the outermost (gas-rich) regions of the disk
become the outermost regions of the resulting tidal tails (TT72), gas
rich extensions to optical tails are to be expected.  The higher
dispersion of outer disk stars with respect to the gas (van der Kruit
1988) and the finite lifetimes of the most luminous stars will
exacerbate this effect, leading to further increases in \MhLb\ in the
outer regions (HvG96). 

However, Arp 299 is quite different from the above mentioned systems
in that it is not just the outermost regions that are gas rich --- the
gas-rich filament appears to extend all the way back to the main body,
running parallel to the optical tail. That is, it appears to cover a
very broad range of radii. So the first question we wish to address is
whether such a feature can arise due to combination of dynamical
and projection effects which cause gas-rich outer regions to be
projected adjacent to optically brighter inner regions.

We believe that answer to this question lies in the dynamical
development of tidal tails, as demonstrated so vividly by TT72.  These
authors emphasized that tails are not linear structures, but actually
two dimensional ``ribbons" twisting through space.  In particular,
Fig.~2 of TT72 shows that the outermost edges of the ribbon come from
the outermost radii of the original disk, while middle regions come
from intermediate radii, etc.; hence, the outer disk material forms a
``sheath" around the inner disk material along its entire length.
Therefore it is possible that a lateral twist may cause the outer edge
of the ribbon to lie in a different plane from the inner regions,
i.e.~for the most gas-rich regions to appear in a different plane from
the less gas-rich, optically brighter regions.

The models also show that the material from different radii form continuous
structures in space and velocity, and hence that the resulting tidal
features will necessarily be continuous in space and velocity as well
(Barnes 1988, Hibbard \& Mihos 1995).  But while the parallel filaments
of Arp 299 appear to be kinematically continuous, they appear
morphologically distinct, or bifurcated.  We suggest that the
bifurcation arises due to a pre-existing warp in the outer regions of
one of the disk.  This suggestion comes from our recent attempts to
match the tidal morphology and kinematics of the northern tail using
N-body simulations (Hibbard \et\ in preparation).  In our preliminary
trials (using only 4096 particles per galaxy), we can rather easily
match either the morphology and kinematics of the outer tidal filament
and N clump, or the kinematics of the inner filament and morphology of
the stellar tail.  However we have been unable to match both features
simultaneously with a single disk orientation and viewing angle.  This
leads us to suspect that the disk of NGC 3690 was initially warped (as
is often the case for disk galaxies; Bosma 1991).  In this case, the
outer, optically fainter regions will be pulled off along one plane
(resulting in the outer tidal filament), while the regions from smaller
radii will move primarily in a different plane (resulting in the inner
filament).  A similar explanation may account for the less extreme
bifurcated structures seen in other tidal tails (e.g.~M81 Yun \et\ 1994;
NGC 3921 HvG96; NGC 2535/6 Kaufman \et\ 1997; NGC 4038 Hibbard, van der
Hulst \& Barnes, in preparation).

However, this projection effect does not explain the low {\hi} content
of the optical tail in Arp 299.  Although the less extreme radii of
spirals have a lower \MhLb\ than the outer regions, they still have
significant quantities of {\hi}.  We therefore seek a separate
explanation for this feature.

\subsubsection{Gas-Poor Inner Filament \label{sec:gaspoor}}

For the original disk to simultaneously give rise to a gas-rich outer
filament and a gas-poor optical filament would require a special {\hi}
radial distribution --- one in which the gaseous disk had a low column
density within the stellar disk, but was very gas-rich beyond this. 
Such distributions are indeed seen in some early type galaxies,
especially SB0's (van Driel \& van Woerden 1991).  However, two facts
lead us to suspect that this is not a viable explanation for Arp 299. 
The first is that such systems usually have an order of magnitude less
{\hi} than observed in the tidal tail of Arp 299 (which itself is less
than half of the original gas content).  The second is that the
starlight in such systems is dominated by an extended central bulge,
which the NIR observations appear to rule out for NGC 3690 (\S
\ref{sec:encounter}). 

Another factor that leads us to consider other explanations is that
Arp 299 is one of a number of peculiar starburst galaxies exhibiting
curious displacements between tidal {\hi} and starlight.  Further, the
five systems which exhibit the most dramatic displacements (M82, Yun
\et\ 1993; NGC 4631/4656, Weliachew \et\ 1978; NGC 520, HvG96; Arp 220
Yun, Hibbard \& Scoville in preparation, HVY99; and Arp 299) also host
massive nuclear starbursts with associated powerful outflows that
extend many kpc from the nuclear regions (``superwinds", e.g.~Heckman,
Lehnert \& Armus 1993; Lehnert \& Heckman 1996), and in each case the
tidal {\hi} shows a displacement or anti-correlation along the
direction of the expanding hot superwind fluid.  This leads us to
suspect a possible interaction between the expanding wind fluid and
the tidal debris (see also Chevalier \& Clegg 1985; Yun \et\ 1993;
Heckman, Lehnert \& Armus 1993; Dahlem \et\ 1996).

This issue is covered in more detail in HVY99, and here we just discuss
the specific details of Arp 299.  In this system the evidence for an
expanding superwind comes from the morphology and kinematics of the
optical emission line gas (Armus, Heckman \& Miley 1990) and the X-ray
observations of Heckman \et\ (1998).  In the latter, the soft X-ray
emission from hot gas is oriented perpendicular to the harder X-ray
emission from discrete x-ray sources associated with the starburst.  The
hot gas emission is elongated roughly along the minor axis of IC 694,
extending to a radius of order 25 kpc (Heckman \et\ 1998).  Since the 
observed X-ray emission represents just the hottest, densest portions 
of the adiabatically expanding wind (Wang 1995), the full extent of the
wind may extend well beyond this. 

The mechanical effect of such a wind may be evidenced in the increased
column density and velocity dispersion of the {\hi} knot in the inner
filament ($r\approx$ 60 kpc), which might be due to the winds
ram-pressure effects.  The wind might also account for the
high-velocity envelope of gas seen in Fig.~\ref{fig:filamentLV} and
the high velocity dispersion regions seen on the southern end of the N
clump in Fig.~\ref{fig:TailOpt}d. However, we do not believe that the
winds ram pressure has pushed the gas fully out of the inner filament
and into the outer filament. Such action should significantly change
the kinematics of the gas, whereas the kinematics of the gas within
both filaments is nearly identical.  Instead, we suspect that the
superwind affects the ionization state of the gas in one of two ways:
either directly by shock ionization, or indirectly by clearing a
sight-line from the tidal tail to the nuclear starburst, allowing UV
photons from the young stars to escape these regions and reach parts
of the tail. Calculations carried out in HVY99 suggests that both
scenarios are in principle capable of affecting gas at the observed
column densities and distances, although there is insufficient
evidence to discriminate between them.

Whether the ``missing'' {\hi} has been ionized or not is subject to
observational investigation.  The expected emission measure in
convienient units is $EM = 0.42 \,{\rm cm^{-6} \, pc}\, ({N_{HI} \over
2\times 10^{20} \, {\rm cm}^{-2}})^2 ({10 \, {\rm kpc} \over dL})^2$,
where $N_{HI}$ is the missing column density of {\hi}, and $dL$ is the
line-of-sight depth of the {\hi} column.  If the gas has a clumpier
distribution, then there should be some higher density peaks with
emission measures well above these levels.  These emmission measures
are within the capabilities of modern CCD detectors (e.g.  Donahue
\et\ 1995 obtained 0.3 cm$^{-6}$ pc in NGC 4631), and such emission
would be well worth looking for.

\subsubsection{Gas/Light Anti-correlation in the N Clump}

The main problem with the above scenarios is that neither can explain
the {\hi} and optical anti-correlation at the end of Arp 299 tail,
whereby there is a local minima in the {\hi} column density where the
optical tail crosses the N Clump, and local maxima to either side of
this (Fig.~\ref{fig:TailOpt}b).  The expanding galactic wind or
photoionization cone should have no idea where the optical tail lies.
In HVY99, we suggest that the outer {\hi} from the warped disk is
already in a highly ionized phase, due to its low density and the
intergalactic UV field.  In this environment, we suggest that the
additional ionizing radiation due to evolved sources within the
stellar tail itself, such as late B stars and/or white dwarfs,
increases the ionization fraction of the local gas over that of the
tidal gas without accompanying starlight. The lower neutral fraction
is then seen as a drop in the neutral gas column density.  A simple
calculation is carried out in HVY99 that suggests that this process is
in principle feasible, if the gas in the N clump has a line-of-sight
thickness \simgt 25 kpc.  In thise case, the expected emission measure
(\simlt 0.02 cm$^{-6}$ pc) is well below what is observable with
current technology.

\subsubsection{Outer morphology: conclusions}

These explanations are somewhat unsatisfactory since we need three
separate mechanisms to explain the three puzzling characteristics: a
warped {\hi} disk to explain the parallel tidal filaments; a starburst
wind or unusual initial {\hi} distribution to explain the lack of
{\hi} within the optical filament; and an increased ionization
fraction in the gas due to the presence of the stellar tail to explain
the anti-correlation between the {\hi} and optical light in the N
clump.  Whatever their cause, these observations point to some
important differences occurring between gas and stars during the
formation of some tidal tails.

\subsection{Conditions for Luminous Infrared Phase \label{sec:IRphase}}

A main objective of our studies on the atomic gas in IR luminous
galaxies ($L_{IR}>3\times 10^{11} L_\odot$, Hibbard \& Yun 1996 and in
preparation) and the molecular gas in the less-IR luminous systems of
the Toomre Sequence (Yun \& Hibbard 1999) is to investigate the
conditions necessary for fueling a highly elevated period of star
formation.  Certainly a high gas content is one of the pre-requisites
for fueling such large starbursts (Sanders \et\ 1988).  We are
interested in knowing whether all mergers between gas rich galaxies
pass through such a phase, or if there are requirements on the
progenitor properties or encounter characteristics to trigger such
bursts.  To gain some insight into how this efficient mode of star
formation is triggered in Arp 299, we draw upon statistical studies of
luminous IR galaxies and the results of numerical simulations and to
interpret our observations.

The tight correlation between molecular gas column density
($\Sigma_{H2}$) and $L_{IR}$, especially when compared with the much
looser correlation of gas mass with $L_{IR}$, shows that IR luminous
starbursts are always precipitated by a corresponding increase in the
nuclear column densities of molecular gas (Scoville \et\ 1994; Solomon
\et\ 1997; Kennicutt 1998; Taniguchi \& Ohyama 1998; Yun \& Hibbard
1999).  The CO observations of Arp 299 (Sargent \& Scoville 1991,
Aalto \et\ 1997, Casoli \et\ 1999) show that the highest
column densities in Arp 299 occur at the nucleus of IC 694, with
$\Sigma_{H2}$ over three times higher than at the nucleus of NGC 3690.
Additionally, HCN mapping observations (Aalto \et\ 1997, Casoli
\et\ 1999) shows that the vast majority of the very dense gas 
($n>10^4 \,{\rm cm}^{-3}$) is associated with the nucleus of IC 694. 
Since the two progenitor galaxies appear to be of nearly equal mass
and of similar total gas content (Table~\ref{tab:progen}), we suggest
that the reason there has been so much more gaseous dissipation
towards the nucleus of IC 694 is due to its retrograde spin geometry
(see also Sanders \et\ 1988).  In such encounters, the gas within the
disk is not pulled into tidal bridges and tails, but rather remains in
the disk where it feels the perturbation from the second system
alternately pulling it outward and pushing it inward during an orbit
(e.g.~TT72; White 1979; Noguchi 1991).  This excites
large epicyclic motions in the gas, which in turn leads to higher
collision rates (e.g. Olson \& Kwan 1990; Noguchi 1991) and hence to
more gaseous dissipation. 

This suggestion would seem to be at odds with the results from
numerical simulations, which find only a moderate dependence of
gaseous dissipation and/or starformation activity on the encounter
spin geometry (e.g. Mihos \& Hernquist 1996; Barnes \& Hernquist
1996). Further, these simulations invariably have the starbursts
occurring only when the nuclei finally coalesce, while in Arp 299 (
and other IR luminous mergers, Murphy \et\ 1996, Mihos \& Bothun 1998),
the nuclei are still well separated ($d>4.7$ kpc). A possible
solution to these concerns is found in the simulations of Noguchi
(1991), who stresses the episodic nature of starbursts in mergers (see
also Olson \& Kwan 1990). Noguchi notes that while the maximum
enhancement of activity is not strongly dependent on the spin geometry
of the encounter, the time evolution of the activity is. In
particular, his prograde-retrograde encounter produces a longer period
of enhanced activity over other spin combinations and also leads to a
slower coalescence of the nuclei.  As a result, the cloud collision
rate is still significantly increased as the nuclei separate (see his
Fig.~9). This model predicts that the most luminous phase of Arp 299's
starburst is still in the future, in which case this merger is
predicted to out-shine even Arp 220.

However, the numerical simulations are necessarily ad-hoc in nature,
and the micro-physics of hydrodynamics (e.g. SPH as in Mihos \&
Hernquist and Barnes \& Hernquist vs. sticky particle as in Noguchi),
star formation, and ``feedback" (energy input back into the
interstellar medium due to star formation) is poorly understood.
Therefore, before drawing too many inference from the simulations, it
would be prudent to first test the numerical formalisms for star
formation, dissipation, and feedback by direct comparison with real
systems.  Until simulations can reproduce starbursts occurring at the
times and locations as observed, we must be cautious when
extrapolating their results too far into the future.

The {\it observational} fact that systems can experience much of their
star formation when the nuclei are well separated has important
ramifications for the structure of the resulting remnant.  In
particular, stars are dissipationless and cannot radiate their orbital
energy away.  Therefore stars that form while the nuclei are well
separated will be spread over a larger range of radii in the resulting
remnant than those that form when the nuclei are practically merged.
This population can experience violent relaxation (e.g. Barnes 1992;
Barnes \& Hernquist 1996), resulting in an extended dynamically hot
population.  The resulting remnant will have a much more regular
luminosity profile than a merger in which the gas all sinks to the
very center of the potential before forming stars (\cf\ concerns
raised in Mihos \& Hernquist 1994).

\subsection{Lack of Tidal Dwarfs \label{sec:Tidwf}}

The widespread lack of small-scale correlation between optical and
gaseous density enhancements within the tidal tail was unexpected.  In
previous {\hi} and optical observations of six double-tailed merging
systems (van der Hulst 1979; Hibbard \& van Gorkom 1993, 1996; Hibbard
\et\ 1994), five dwarf galaxy like concentrations of gas, light and
\Ha\ emission appear entrained within tidal tails {\it and} coincident
with distinct features in the optical color maps and the {\hi}
velocity dispersion maps.  These concentrations have been interpreted
as ``tidal dwarf galaxies" forming out of the expanding tidal material
(see also Schweitzer 1978; Mirabel, Dottori \& Lutz 1992; Duc \et\
1997; Malphrus \et\ 1997), an idea that has been supported by
numerical simulations (Barnes \& Hernquist 1992, 1996; Elmegreen,
Kaufmann \& Thomasson 1992; Hibbard \& Mihos 1995).  In addition to
these dwarf-sized concentrations, the previously observed tidal tails
often contain numerous knots of gas, light, and star-forming regions
(see also Hutchings 1996; Hunsberger, Charlton \& Zaritski 1996; Mihos
\& Bothun 1998).  We therefore fully expected similar features to
appear in a tidal feature as long and as massive as that seen in Arp
299.

However, the optical tail of Arp 299 is remarkably smooth and
featureless (Fig.~\ref{fig:DeepOpt}).  There are a number of point
sources falling upon the optical tail, but the surface density of such
points is similar to that of point source elsewhere in the field.
Their colors are generally quite red ($B-R>1.7$,
Fig.~\ref{fig:TailOpt}a), and most are probably background galaxies.
And while there is considerable structure within the {\hi} tail, there
are no {\hi} peaks with a corresponding enhancement in the underlying
starlight (Fig.~\ref{fig:TailOpt}b).  Further, there are only three
regions which exhibits a significant ($>$20\%) increase in the {\hi}
velocity linewidth (indicated in Fig.~\ref{fig:TailOpt}d) --- along
the southern edge of the N clump and at the {\hi} knot in the inner
tidal filament (\S \ref{sec:outerKin}) --- and these features do not
correspond to any notable features in the optical images or color
maps.

To further investigate this question, we estimate the importance of
self-gravity within the tail by comparing the luminous mass
($M_{lum}$) to the dynamical mass ($M_{dyn}$).  The luminous mass is
estimated by adding the gas mass and the stellar mass.
The latter is estimated from the \B-band surface brightness and adopting a
stellar mass to light ratio of $(M/L)_B=2 M_\odot L_\odot^{-1}$,
characteristic of stellar disks (Bottema 1993). The dynamical mass is
calculated via $M_{dyn}=1.76\times 10^6 M_\odot \times \sigma_{HI}^2
\times r_{1/2}$ (Binney \& Tremain 1987; see also Kaufman \et\ 1997),
where $r_{1/2}$ is the FWHM radius of the beam ($\sim$ 1.9 kpc for
the high resolution data). This will over-estimate the effects of
self-gravity if (1) the line-of-sight dimension of the tail is larger
than the beam width ($\sim$ 3.8 kpc), in which case the above
prescription overestimates the stellar mass; and if (2) the stars
contribute significantly to $M_{lum}$ and have a velocity dispersion
greater than $\sigma_{HI}$ (e.g.~van der Kruit 1988).  Since we
suspect both of these conditions to hold, our calculation of
$M_{lum}/M_{dyn}$ should be a stringent upper-limit. We derive a
lower limit to $M_{lum}/M_{dyn}$ by setting $M_{lum} = M_{gas}$.

Using the high resolution data, we find that $M_{lum}/M_{dyn} < 0.3$
everywhere, with the highest values on the brighter optical peaks of
the inner tidal filament\footnote{Using a velocity dispersion of 20
\kms\ as inferred for the stars in \S\ref{sec:outerMorph} would give 
$M_{lum}/M_{dyn} <$ 0.04 on this feature}.  The peaks in the {\hi}
column density along the outer filament and in the N clump have $0.2 <
M_{lum}/M_{dyn} < 0.3$, and the {\hi} knot in the inner filament has
$0.1 < M_{lum}/M_{dyn} < 0.2$.  For comparison, the tidal dwarfs in
the tail of NGC 7252 (Hibbard \et\ 1994), NGC 3921 (Hibbard \& van
Gorkom 1993, HvG96) and NGC 4038/9 (Mirabel, Dottori \& Lutz 1992)
have 0.50 \simlt $M_{lum}/M_{dyn}$ \simlt 1.  Therefore Arp 299 lacks
the potentially self-gravitating dwarf-like concentrations seen in
other tailed systems, and we conclude that tidal dwarf formation is
not a ubiquitous process.

This suggests that the formation of such sub-structures may depend on
factors other than the ability to raise a decent sized tail.  We note
that other tidal tails which contain candidate tidal dwarf galaxies
have similar gas column densities as in the tail of Arp 299, but have
optical tails that are \about 2 \msqas\ brighter.  Additionally, in
these systems the gas and optical material coincide.  Both of these
factors increasing the underlying mass density within the tail.
Further, since the gas in observed tidal dwarfs are a significant
contribution to the dynamical mass ($M_{gas}/M_{dyn}\approx 0.5$), it
may be the case that whatever process removes the {\hi} from the inner
tidal filament (see \S \ref{sec:gaspoor}) also suspends the dwarf
formation process.

Whatever the explanation, the present observations suggest that one
should be cautions about simply measuring the mass or light contained
within the highest gas column density contours or optical isophotes and
calling these concentrations distinct entities. 

\section{Conclusions}
\label{sec:concl}

\begin{itemize} \item{A new tidal tail has been discovered in the
on-going merger Arp 299.  The tail stretches 180 kpc to the north, and
must have taken at least 750 Myr to form.  This is much longer than
the age of the current starburst, suggesting that the starburst and
merger clocks start at different times and run at different rates. If
the present starburst began when the tidal tail was first launched,
the spread of ages in the resulting remnant will be much larger than
is usually assumed.}

\item{What had previously been interpreted as a single rotating gaseous
disk surrounding the two merging systems is shown to be a gaseous disk
associated with the eastern system, IC~694.  
The kinematic information in concert with the optical morphology
allows us to deduce that NGC~3690 experienced a prograde encounter
while IC~694 experienced a retrograde encounter.  From the cold gas
contents and NIR morphologies, we suggest late type progenitors for
both systems (Sab-Sb for IC 694, Sbc-Sc for NGC 3690).}

\item{The observations show that Arp 299 is experiencing an IR bright
starbursts while the nuclei are still well separated ($dr$\simgt 4.7
kpc). We suggest that this is due in part to the retrograde spin
geometry of IC 694.}

\item{The tidal morphology is unique among mergers: there is a low
surface brightness stellar tail with low column density {\hi} and a
conspicuous parallel gaseous tail of higher {\hi} column density and
with little, if any, associated starlight.  The tails have very similar
kinematics suggesting that they are physically connected.  We suggest
that such bifurcated or parallel tails result from a progenitor with a
warped disk.  However, this explanation does not explain the low {\hi}
content of the optical tail.}

\item{The similarity between this morphology and that of other starburst
galaxies with direct evidence of superwind outflows (M82, NGC 4631,
NGC 520, Arp 220) suggest that the galactic superwind recently
imaged by Heckman \et\ (1998) may be responsible for clearing the
optical tail of much of its cold diffuse gas. 
Local ionizing sources within the tidal tails may also play a role
in some of the observed optical/gaseous anti-correlation. These
issues are explored in more detail in HVY99}

\item{The absence of dwarf galaxy-like mass concentrations in the tidal tail
suggests that tidal dwarf formation is not a ubiquitous process.  It
is possible that the whatever process removed gas from the optical
tail may have consequently suppressed the dwarf formation process.}

\end{itemize}

\bigskip

\acknowledgments

We would like to thank S.~Aalto, L.~Armus, F.~Casoli \& F.~Combes for
sharing of data and information prior to publication.  We thank the
referee, Tim Heckman, for a timely and useful referees report which
helped clarify many areas of this paper. JEH would like to thank the
Owens Valley group at Caltech, and in particular Nick Scoville, for
the hospitality extended during the summer of 1995 to reduce the {\hi}
data presented in this paper.  Thanks also to W.~Vacca, J.~van Gorkom,
J.~Barnes and G.~Stinson for useful conversations.  JEH acknowledges
support by Grant HF--1059.01--94A from the Space Telescope Science
Institute, which is operated by the Association of Universities for
Research in Astronomy, Inc., under NASA contract NAS5--26555, for much
of this research.

Most of the figures in this paper were made with the WIP interactive
software package (Morgan 1995). This research has made extensive use
of the NASA/ IPAC Extragalactic Database (NED), which is operated by
the Jet Propulsion Laboratory, California Institute of Technology,
under contract with the National Aeronautics and Space Administration.
The Digitized Sky Surveys were produced at the Space Telescope Science
Institute under U.S.~Government grant NAG W-2166.  The images of these
surveys are based on photographic data obtained using the Oschin
Schmidt Telescope on Palomar Mountain and the UK Schmidt Telescope.
The plates were processed into the present compressed digital form
with the permission of these institutions.

\begin{appendix}
\section{Who is IC 694?}
\label{sec:append}

There is some confusion in the printed and electronic literature as to
which systems actually comprise Arp 299 (e.g.~IAU circular 6859;
NED)\footnote{For this discussion, it is instructive to look at an
earlier photographic representation of the Arp 299.  For this, we
refer to reader to Plate 1 of Morgan (1958).  In this figure, the NW
spheroidal appears as a very faint fuzz well separated from and much
dimmer than either of the two interacting disks, which appear as
distinct objects.}.  Arp does not offer many clues, and even confuses
the matter by incorrectly designating Arp 296 as NGC 3690+IC 694 in
Tables~1 \& 2 of his Atlas, while giving Arp 299 no further
designation (Arp 1966). However the description in Arp's Table~1, and
the order of his figures (Arp 296 appearing next to other bridge-tail
systems) makes it clear that Arp published the pictures under the
correct number, and gave them the correct description in Table 1, and
just got the ``Designation" switched in both tables.

\begin{figure*}
\epsscale{1.0}
\plotone{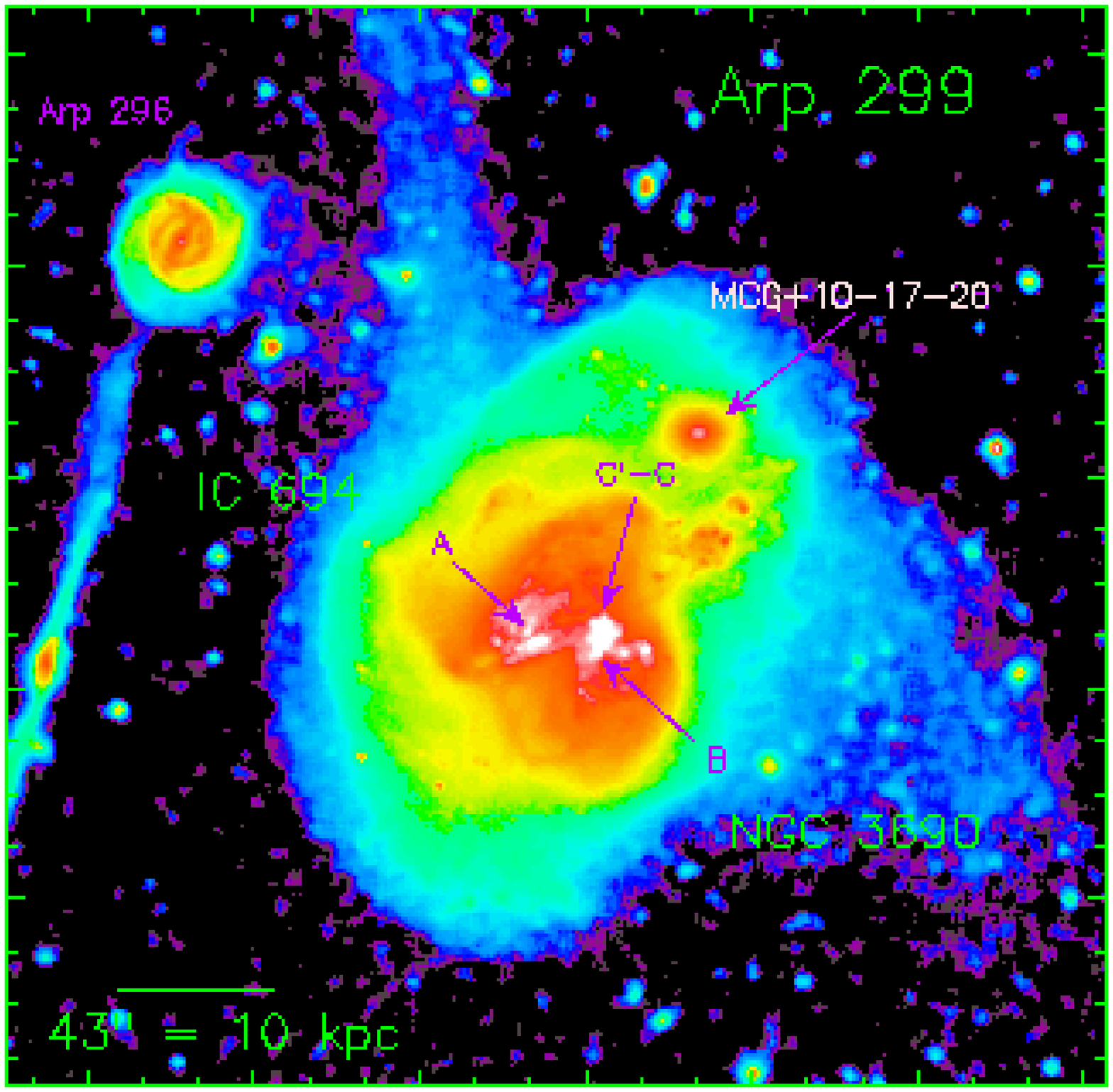}
\caption{\B-band image of the Arp 299 system.
The optical image is displayed with a logarithmic transfer function, with
white set to 28.5 \msqas\ and black set to 22.75 \msqas.  IC 694 is to
the left (east) and NGC 3690 is to the right, and north is up.
Regions denoted as A,B, and C-C$^\prime$ by Gehrz, Sramek \& Weedman (1983)
are so labelled.  We also identify the associated system MCG+10-17-2a
and the unrelated background bar/bridge/tail system Arp 296.  A scale
bar is drawn indicating 10 kpc, assuming a distance of 48 Mpc to Arp
299.  The putative nuclei of the progenitors, as judge by NIR imaging
and spectroscopy, are at regions A and B.  Both of these locations are 
the sites of very dense molecular gas concentrations.  The label C-C$^\prime$
indicates the region of disk overlap, which is also the site of a 
significant concentration of molecular gas. 
\label{fig:a299B}}
\end{figure*}

The description given for NGC 3690 in the NGC
(Dreyer 1888)
translates as
``pretty bright, pretty small, very little extended at PA=80, pretty
gradually brighter towards the middle, small stars south following
near".  According to the NGC ``pretty small" means a diameter of
50\as-60\as, and this radius would incorporate much of the brighter
parts of the system (including the nucleus of the eastern system).  
Given that the five highest surface brightness objects are within 
7\as\ of source {\bf B} in the western system, through a small 
telescope this object must have looked like a single nucleus at 
the location of source {\bf B} in Fig.~\ref{fig:a299B} with a cloud
of surrounding nebulosity, and it is unlikely that two nuclei could
have been discerned.  It therefore seems likely that the original
designation refers to the entire Arp 299 system, although only one
nucleus was identified (corresponding to the western system, what we 
have been calling NGC 3690 in the above).  Therefore whether NGC 3690 
should strictly apply to the entire system or to only the western system
depends on if you believe galaxies are identified by their nuclei or
their integrated light.  We remain agnostic on this point, and from here
investigate which object was designated as IC 694. 

The earliest and clearest association of IC 694 with the eastern
system can be traced to Vorontsov-Vel'yaminov, in his {\it Atlas and
Catalogue of Interacting Galaxies} (Vorontsov-Vel'yaminov 1959), where
Arp 299 appears as VV118a-e.  Source VV118a is the galaxy to the East
(what we have been referring to as IC 694), VV118b the system to the
west (what we have been calling NGC 3690), VV118c is MCG+10-17-2a, the
compact spheroidal 1$^\prime$ to the NW, and VV118d \& e are bright HII
regions in the outer disk regions to the NW.  In the accompanying
table to the Atlas, VV118a is identified as IC 694, VV118b is identified 
as NGC 3690, and sources c,d or e have no other designations.  The vast
majority of researchers since have followed this designation.
Nilson first brings up the question of 
the identification of IC 694 with VV 118a in the UGC (Nilson 1973), 
in his notes on UGC 6471/2:
``Arp 296 (identified as NGC 3690 + IC 694 by Arp) are faint objects
north-preceding UGC 06471 + UGC 06472, magnitude approximately 17 and
21 identification of IC 694 uncertain, may be a small object
north-preceding and inside the outer parts of the double system".
[The use of the name Arp 296 is immediately attributed to Arp's error
in making his tables, discussed above].

IC 694 was discovered by Swift (1893), whose discovery description
translates as ``close double
with NGC 3690 = object 247 of list I of Sir William Herschel.  Suspected
at 132x magnification, verified at 200x".  In the {\it Index Catalogue
of Nebulae} (IC, Dreyer 1895) this becomes 
``very small, forms double nebula with no.~247 of list I of Sir
William Herschel".  ``very small" translates as 10\as-20\as\
diameter, which is on the large side for the NW spheroidal
(\FWHM\about4\as).  There is no definition of the term ``double" in
the IC\footnote{Ironically, in the introduction to the IC Dreyer
states ``the system of abbreviated description ...  has been in use so
long that it is unnecessary to enter into a lengthy explanation of
it.."}, and perhaps this is where the confusion lies.  Certainly with
todays knowledge we would call Arp 299E+W a double, and if anything
the spheroidal to the NW as a possible companion.  However, is this
how Dreyer and Swift used the term?

So the question is, when Swift identified IC 694, did he subdivide the
NGC 3690 entry, or add an entry for the faint nebulosity to the NW?
The position for IC 694 is given as 11:20:44 +59:20 (epoch 1860) in
the IC, while the NGC position for NGC 3690 is 11:20:45 +59:19, {\it
i.e.~IC 694 lies approximately 1$^\prime$ to the NW of NGC 3690}.
This seems like pretty conclusive evidence that IC 694 indeed refers
to the spheroidal to the NW.  However, Dreyer remarks in the
introduction to the IC that Swifts positions are ``generally reliable
within one or two minutes of arc, but larger errors occur
occasionally..."  We also learn from Dreyer that Swifts observations
were taken through a 16\as\ refractor at the Rochester Observatory.
Since the NW galaxy is 3 mag fainter than either of the disk galaxies
($m_B$=16.2 mag from CCD observations reported above), is it even
possible for Swift to have seen the NW companion? If not, it is likely
that he simply resolved NGC 3690 into two separate condensations of
nearly equal luminosity, adding a separate catalogue entry for the
second system.  A similar super-classification appears to have
happened in the case of NGC 4861/IC 3961 (Arp 266) and NGC 2207/IC
2163.

Recent attempts to repeat these observations with similar sized
telescopes by amateur astronomers suggest that (1) the second light
concentration now associate with the nucleus of Arp 299E is very
easy to discern as distinct from Arp 299W; and (2) the NW spheroidal is
indeed a {\it very} difficult, but not an impossible, object.  Combined
with the correct relative position of IC 694 given by Swift, it seems
that there is strong evidence that IC 694 properly designates the NW
compact spheroidal.  However, since these observations were made with
prior knowledge of both the existence and location of the NW spheroidal,
this confirmation is not entirely independent. 
Since it is nearly impossible to prove this issue conclusively one way
or the other, we choose to follow the vast majority of researchers and
use the designation introduced by Vorontsov-Vel'yaminov (1959),
concluding as Sulentic \& Tift (1973) that ``Any further question
as to which objects Dreyer was referring to can only be of historical
interest". 

JEH offers special thanks to Harold Corwin, Steve Gottleib, Malcom
Thomson, Bill Vacca and Dennis Webb for many useful discussions on this
issue. 

\end{appendix}

\clearpage

\onecolumn
\def\b{beam$^{-1}$}

\begin{figure*}
\plotone{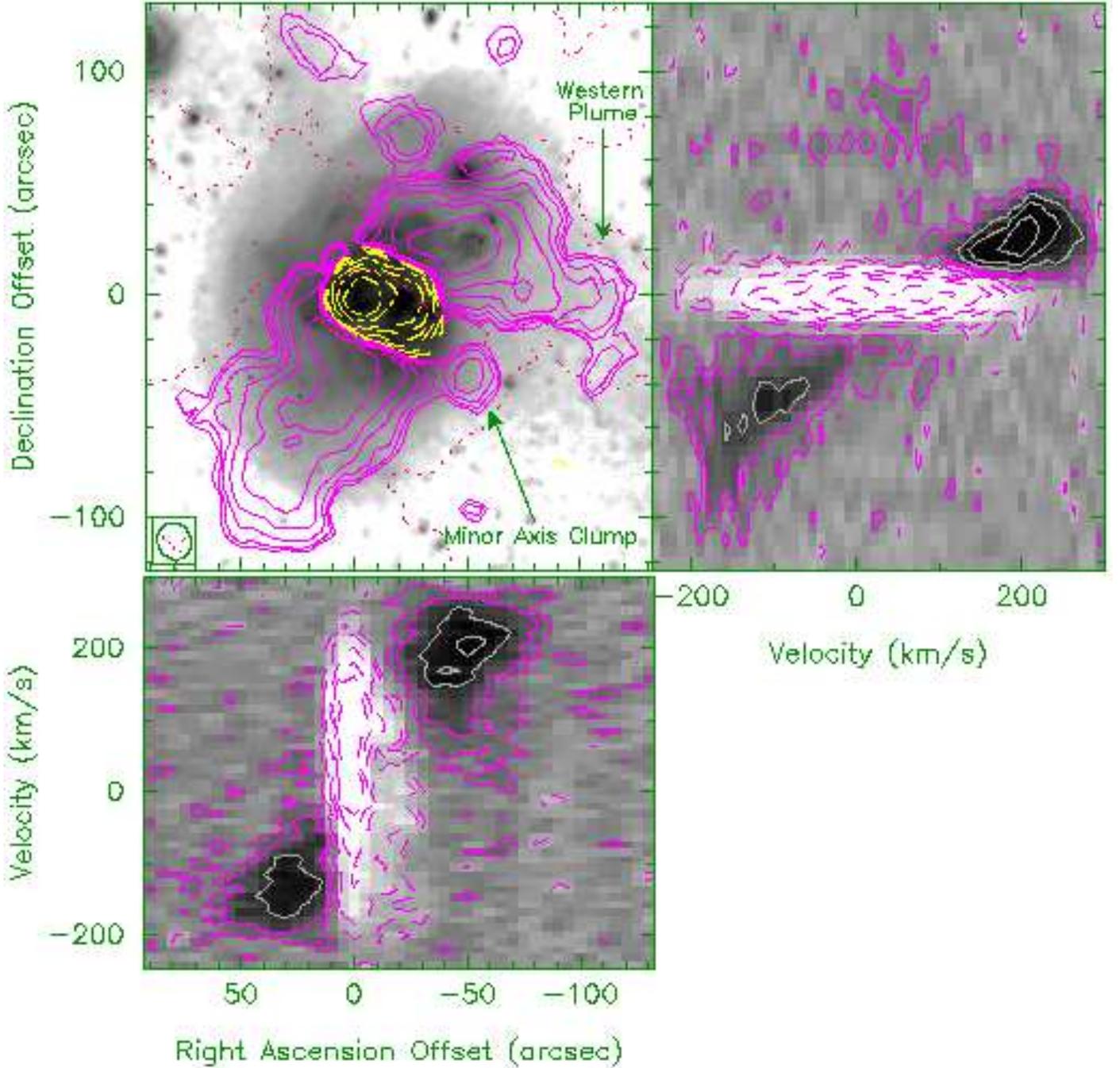}
\caption{Position-Velocity profiles of the inner regions of
Arp 299, demonstrating the kinematics of the inner {\hi} features.
The top left panel presents the \B-band image displayed as in
Fig.~\ref{fig:a299B}, with contours from the high resolution {\hi}
integrated intensity map superimposed (FWHM=17\as\x 15\as, as
indicated by the ellipse in the lower left corner of the panel).
Negative contours are drawn as dotted white lines, indicating the
region of {\hi} absorption.  The positive contours begin at 11.05 mJy
\b\ km s$^{-1}$ (5\col{19}), while negative contours begin at -33.15
mJy \b\ km s$^{-1}$.  Successive contour levels are twice the previous
level.  The black dotted contour indicates a column density of
1\col{19} from the low-resolution \hi\ data (\FWHM=35\as).  To the
right is a position-velocity map of the high resolution data after
summing the emission in right ascension, and below is a similar plot,
made by summing the emission in declination.  Positions are measured
relative to the nucleus of IC 694, and velocities are measured
relative to 3100 \kms.
\label{fig:innerLV}}
\end{figure*}

\begin{figure*}
\plotone{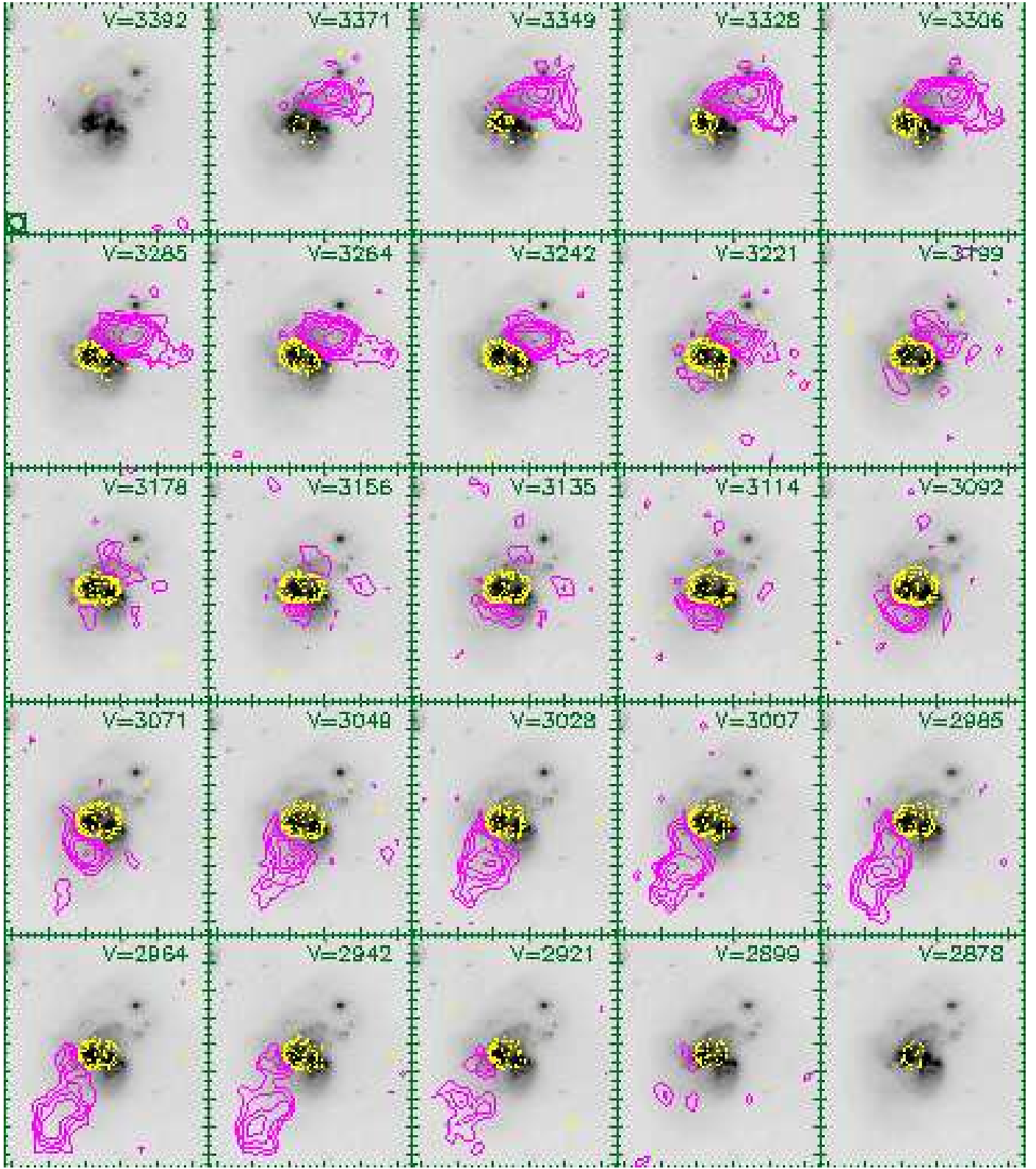}
\caption{\hi\ channel maps of the inner regions of 
Arp 299 contoured upon the \B-band image.  The higher resolution data
are used after Hanning smoothing in velocity by a factor of two to a
channel spacing of 21.4 \kms. The beam size (17\as\x 15\as) is
indicated in the lower left hand corner of the first panel, and each
panel is labelled with the channel number and heliocentric velocity.
Dotted white lines indicate negative contours, which are drawn at
1$\sigma \times$ [-3, -5, -7, -10], where 1$\sigma$=0.25 mJy \b\ is
the single channel noise level.  Positive contours are drawn at
1$\sigma \times$ [3, 5, 7, 10], with higher contours drawn a factor of
1.5 times the previous level.  The lowest contour corresponds to a
column density of 7.2\col{19}.
\label{fig:innerChan}}
\end{figure*}

\begin{figure*}
\plotone{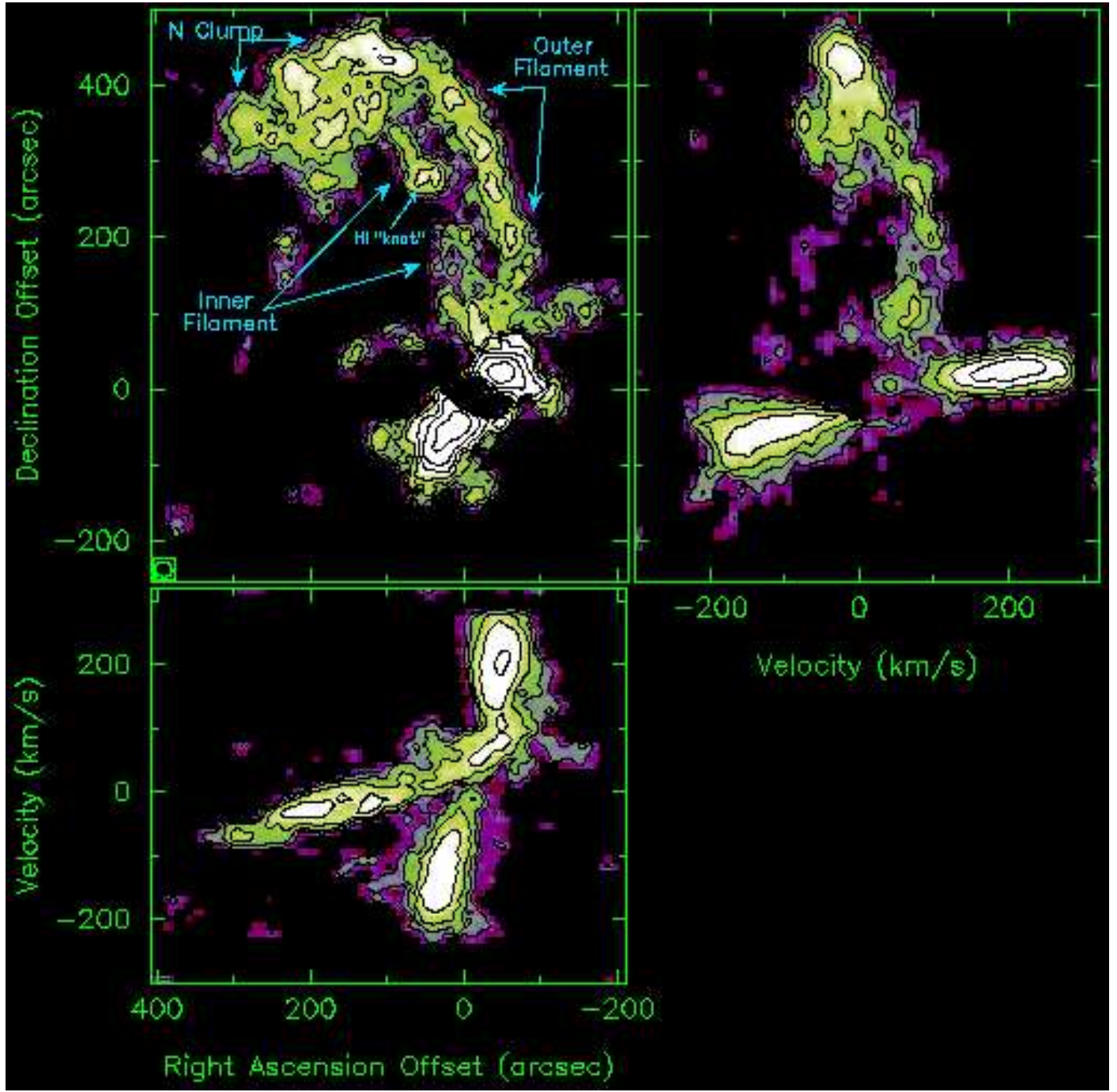}
\caption{Position-Velocity profiles of the entire Arp 299
system, demonstrating the kinematics of the tidal {\hi} features.
The top left panel presents a false color and contoured image of the
intermediate resolution data cube (22\as\x 20\as). Features referred
to in the text are labelled. The adjacent panels are position-velocity
profiles, constructed by summing the data cube in right ascension
(right) or declination (bottom).  Values below 1$\sigma$ were not not
used when constructing the position-velocity plots. Contours are drawn
starting at a level of 8 mJy \b\ km s$^{-1}$ (2\col{19}), with
higher contours drawn a factor of 2 times the previous level.  
\label{fig:outerLV}}
\end{figure*}

\begin{figure*}
\epsscale{0.8}
\caption{Deep optical image of the entire Arp 299 CCD mosaic,
showing the total extent of the stellar tail.  The deep \B- and \R-band data
are shown on the left and right, respectively, using a logarithmic 
transfer function. The upper panels ({\bf a \& b}) show a greyscaled 
representation  of the data, while the lower panels add contours. 
The greyscales give a better feeling for the structure at the lowest 
light levels, while the contoured images provide a quantitative measure 
of the light peaks.  The greyscales cover the range \mB=28.5 \msqas -- 
22.75 \msqas\ and \mR=28.5 \msqas -- 22.0 \msqas.  In the upper panels, 
white contours are drawn every 1 \msqas\ starting at \mB=24 \msqas\ and 
\mR=25 \msqas. In the lower panels black contours are added every 1 
\msqas\ starting at \mB=28 \msqas\ and \mR=27 \msqas. Three stars
referred to in the text are labelled in panel (a). 
\label{fig:DeepOpt}}
\end{figure*}

\begin{figure*}
\caption{An RGB representation of the Arp 299 optical 
and HI data. Yellow and green indicates optical emission, while blue
indicates {\hi} emission. Regions containing both forms of emission
appear as the sum of these colors. This figure emphasizes the 
differences between the stellar and gaseous tidal morphologies. 
\label{fig:a299rgb}}
\end{figure*}

\begin{figure*}
\plotone{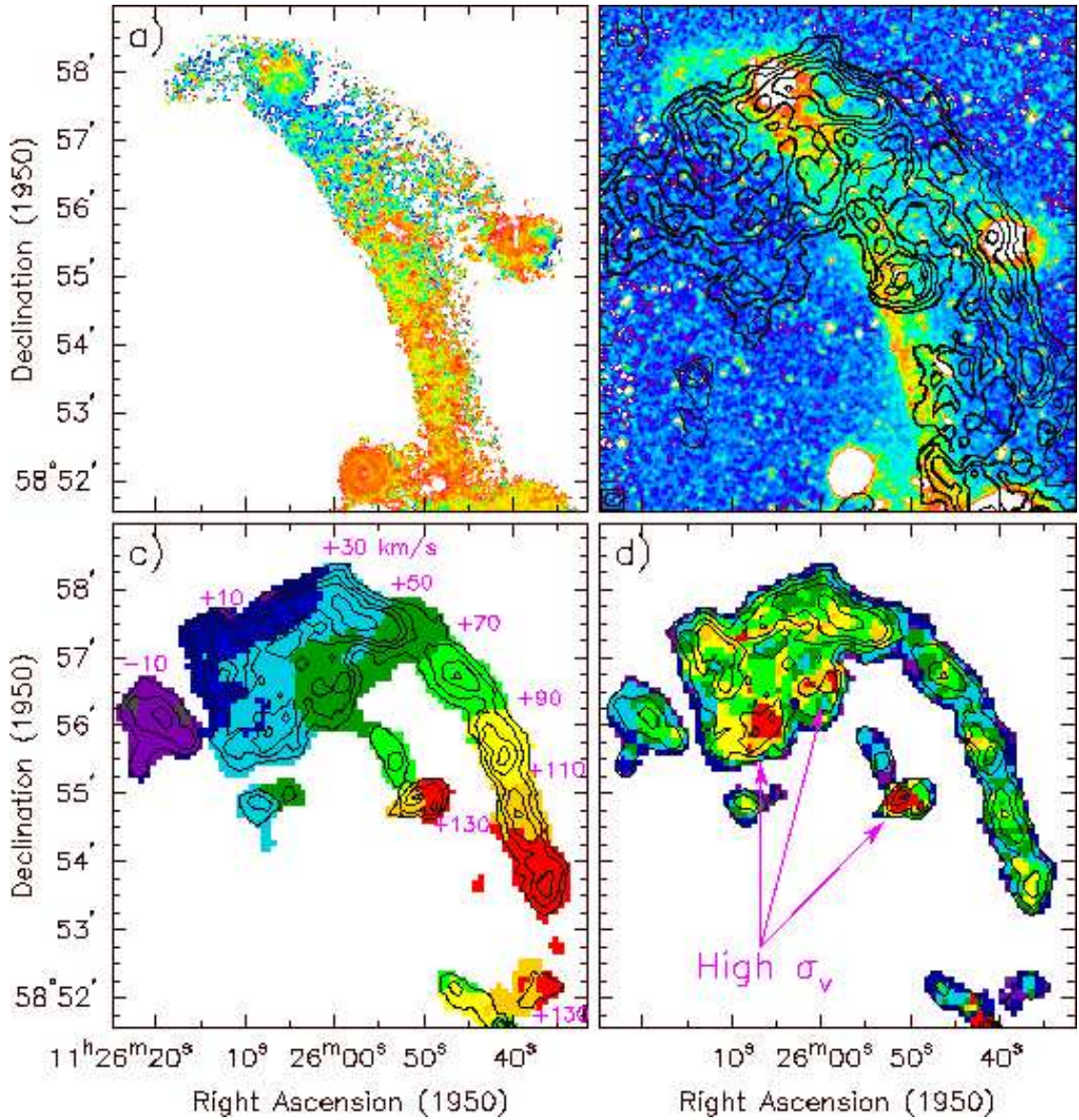}
\caption{Detail of the Arp 299 tidal tail, showing the
relationship between the gaseous and stellar morphologies and
kinematics. {\bf (a)} \B-\R\ color map constructed from the
smoothed data using only pixels with a signal to noise greater than or
equal to 3 in both the \B- and \R-band images. A spectral transfer
function is used with the following color mappings: Purple: $B-R<$-0.2
mag, blues: $B-R$ = -0.2 -- +1.0; green-yellow: $B-R$ = 1.0 -- 2.0
mag; red: $B-R>$2.0 mag. {\bf (b)} False color rendition of the
\B-band data with contours of the intermediate resolution data
overlaid. Contours are drawn at levels of 4 mJy km s$^{-1}$ \b\
(1\col{19}) \x [3, 6 10, 15, 20, 25]. {\bf (c)} intensity weighted
mean {\hi} line-of-sight velocity. The intermediate resolution has
again been used, but with a higher threshold for the moment algorithm. 
The color boundaries are indicated with labels giving the velocity 
(in km s$^{-1}$) with respect to NGC 3690 ($V_{hel}\approx$3040 \kms, 
\S \ref{sec:absorp}).  Contours are the
same as in (b).  {\bf (d)} {\hi} line-of-sight velocity dispersion,
using the same data and threshold as in (c). The colors correspond to
the following velocity dispersion ranges: purple--blue = 5--8 \kms;
green = 8--10 \kms; yellow--orange = 10--13 \kms; red = 13--20
\kms. The regions of high velocity dispersion discussed in \S
\ref{sec:outerKin} and \S \ref{sec:Tidwf} are indicated.
\label{fig:TailOpt}}
\end{figure*}

\begin{figure*}
\epsscale{0.75}
\plotone{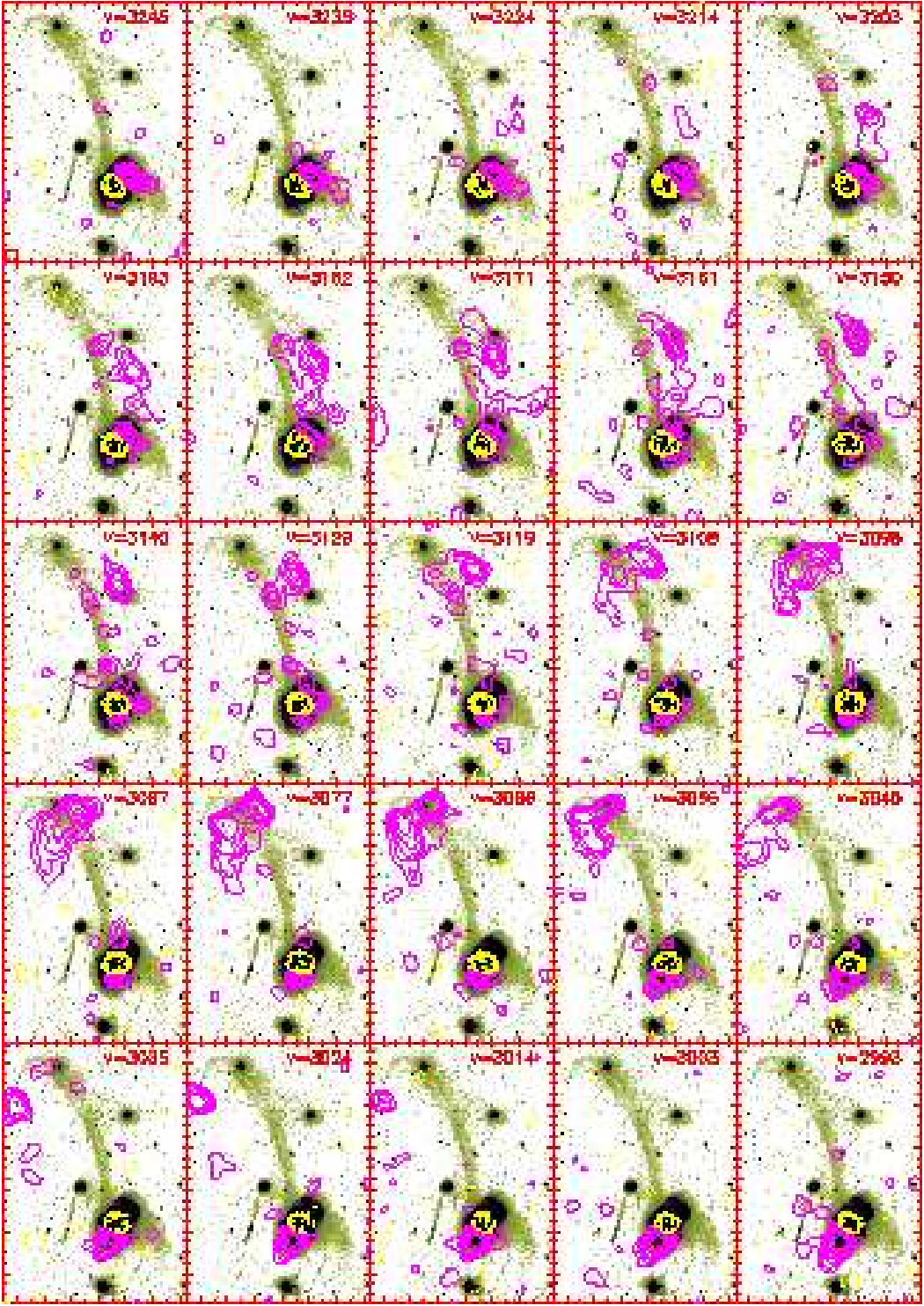}
\caption{Channel maps of the tidal regions of Arp 299 
contoured upon the \B-band image.  The greyscales are drawn from
\mB=28.5 \msqas\ (white) to 25 \msqas\ (black).  The more sensitive
low-resolution {\hi} cube is used at full velocity resolution (10.52
\kms\ channel width).  The 35\as\ beam size is indicated in the lower
left hand corner of the first panel, and each panel is labelled with
the channel number and heliocentric velocity.  Dotted yellow lines
indicate negative contours, which are drawn at 1$\sigma \times$ [-3,
-5, -7, -10], where 1$\sigma$=0.37 mJy \b\ is the single channel noise
level.  Positive contours are drawn at levels of 1$\sigma \times$ [3,
5, 7, 10, 15, 22.5]. The lowest contour corresponds to a column
density of 1\col{19}.  Only channels showing {\hi} emission from the
tail are shown.
\label{fig:outerChan}}
\end{figure*}

\begin{figure*}
\epsscale{1.05}
\plotone{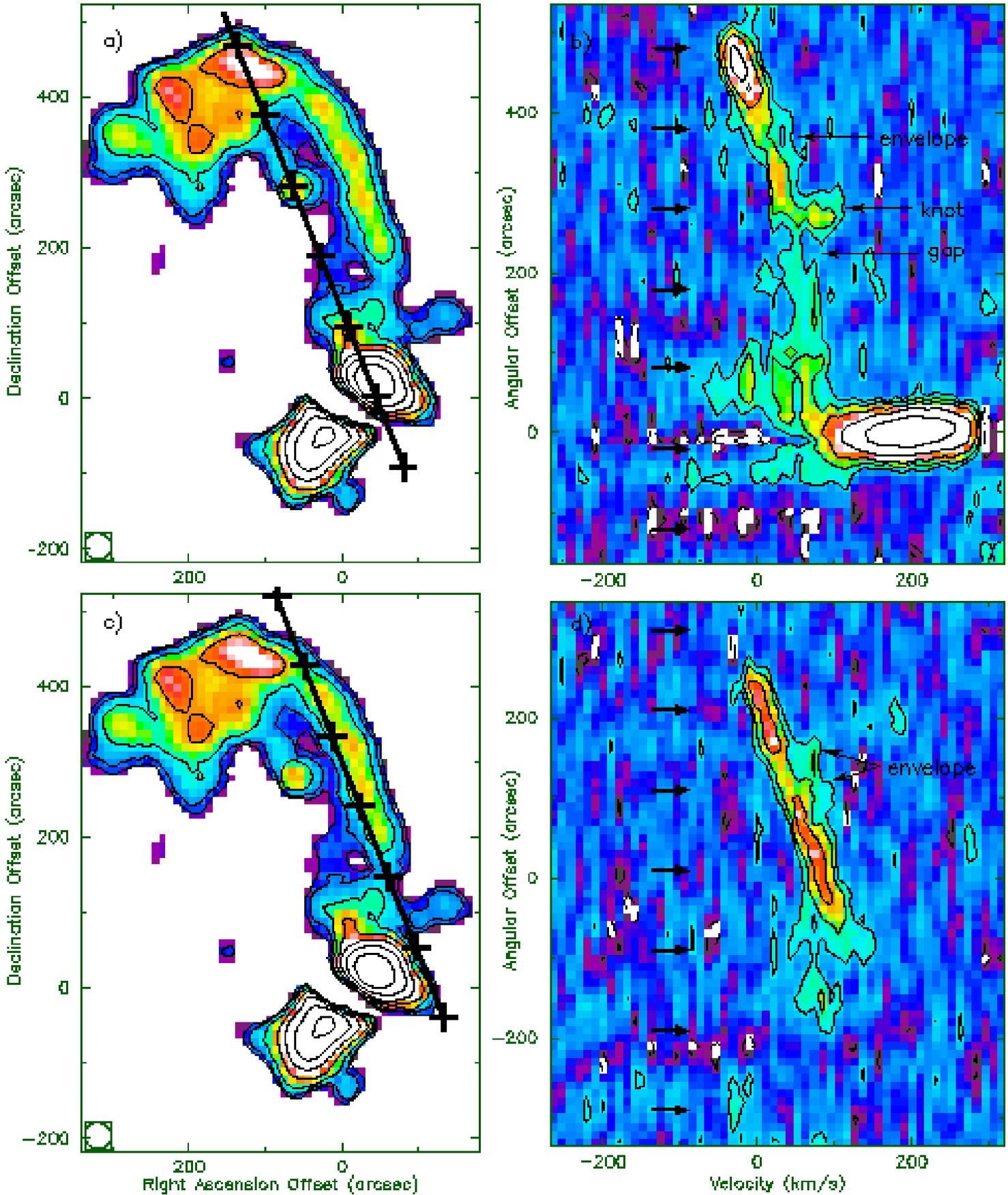}
\caption{Position-Velocity slices taken along 
both gaseous filaments.  The left panels show the location of the
slice on the low-resolution \hi\ iintegrated intensity maps, while the
right panels show the corresponding position-velocity profile. For the
latter, distance along the slice is plotted along the $y-$axis, and
velocity with respect to 3100 \kms\ is plotted along the $x-$axis.
Crosses are drawn at intervals of 100\as\ along the slice in the left
panel, and arrows indicate the corresponding locations in the right
panel. Specific regions referred to in the text are labelled. {\bf (a)
\& (b)} Position-Velocity slice taken along the inner tidal
filament. {\bf (c) \& (d)} a similar slice taken along the
outer tidal filament.
\label{fig:filamentLV}}
\end{figure*}

\begin{figure*}
\plotone{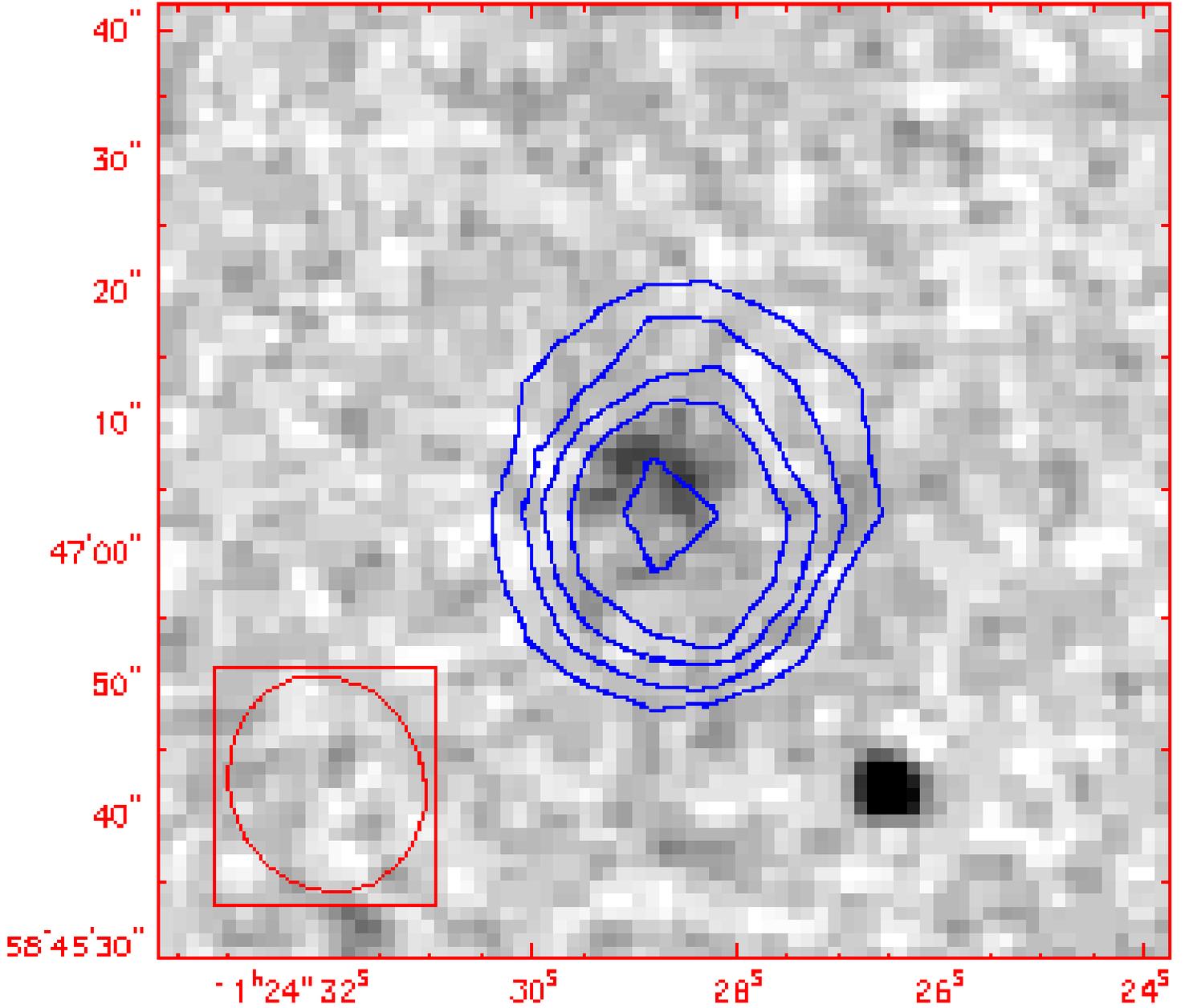}
\caption{Greyscale image of the companion to Arp 299, taken
from Version II of the Digitized Sky Survey, with {\hi} contours
superimposed.  The higher resolution data is used with the 17\as\x
15\as\ beam indicated by the ensquared ellipse to the lower
left. {\hi} column density contours are drawn at [5, 10, 15, 20, 30]
\col{19}.
\label{fig:compMom0}}
\end{figure*}

\begin{figure*}
\caption{\hi\ channel maps of the companion to Arp 299
contoured upon a greyscale of Version II of the Digitized Sky Survey.  
The higher resolution (17\as\x 15\as), full velocity resolution 
($\Delta v$=10.52 \kms) {\hi} data are used, with the spatial 
resolution indicated by the ensquared ellipse to the lower left of 
the first panel.  Contours are drawn at 1$\sigma$ \x [-3, 3, 5, 7], 
where 1$\sigma$=0.20 mJy \b\ is the single channel noise level, 
corresponding to a column density of 9.2\col{18}.  
\label{fig:compChan}}
\end{figure*}
\clearpage

\def\ch{ch$^{-1}$}
\begin{deluxetable}{lrrr}
\tablecaption{VLA Observing Parameters}
\label{tab:HIobs}
\tablehead{
\colhead{Parameter}
}
\startdata
Phase Center ($\alpha_{1950}~~~\delta_{1950}$) & 
$11^h 25^m$ 44\fs 2 & $+58^\circ 50^\prime 18^{\prime\prime}$ \nl
Velocity Center (Heliocentric)&    & 3080 \kms & \nl
Primary Beam (FWHM)&               & 30$^\prime$ & \nl
Phase Calibrator   &               & 1203+645 & \nl
Flux Calibrator    &               & 3C268 & \nl
Bandwidth          &               & 3.125 MHz & \nl
Number of Channels &               & 63 & \nl
Channel Separation &               & 10.5 \kms & \nl\nl
\tableline\nl
Array Configuration                   & D	& C	& C+D \nl
Date                                  & 4/18/95	& 12/10/94 & --- \nl
Synthesized Beam \nl
--- FWHM: Major Axis \x\ Minor Axis  
 & {  56\as \x 46\as } 
 & {  17\as \x 15\as }   
 & {  22\as \x 20\as } \nl
--- Position Angle (East of North)&$+37^\circ$ & $-22^\circ$ & $-12^\circ$\nl
Time on Source (hrs)                  & 2.1     & 2.9     & 5.0 \nl
Noise Level (1$\sigma$) \nl
--- Flux Density (mJy \b)             & 0.56    & 0.37    & 0.30 \nl
--- Column Density (\col{19} \b\ \ch )& 0.25    & 1.7     & 0.80 \nl
--- Brightness Temperature (K \b)     & 0.13    & 0.88    & 0.41 \nl
\enddata
\end{deluxetable}


\begin{deluxetable}{lrr}
\tablecaption{Optical Observing Parameters}
\label{tab:OPTobs}
\tablehead{
\colhead{Parameter}
}
\startdata
Telescope        & & UH 88\as		\nl
Detector         & & Tek 2048		\nl
Readout Mode     & & binned 1\x1 	\nl
Focal Ratio      & & f/10       	\nl
Pixel Size       & & $0\farcs22$	\nl
Field of View:   \nl             
--- Single CCD frame & & $7\farcm5\times 7\farcm5$ \nl
--- Final Image  & & $8\farcm7\times 12\farcm4$     \nl \nl
\tableline\nl
Date             & 6/03/95     & 01/07/96	\nl
Filters          & $R$         & $B$		\nl
Effective Exposure Time  & $\sim 2\times600\,$s & $\sim 2\times900\,$s \nl
Seeing           & $1\farcs2$  & $0\farcs9$	\nl
Sky Brightness ($\rm mag~arcsec^{-2}$) 
		 & 20.9        & 22.9 \nl
$1\sigma$ Sky Noise\tablenotemark{a} ~($\rm mag~arcsec^{-2}$)
                 & 25.8        &  27.0 \nl
$3\sigma$ Sky Noise\tablenotemark{b} ~($\rm mag~arcsec^{-2}$)
                 & 26.9        &  28.0 \nl
\enddata
\tablenotetext{a}{1$\sigma$ noise in original unbinned image}
\tablenotetext{b}{3$\sigma$ noise after binning 9$\times$9; 
this should be close to the 3$\sigma$ detection limit, and we 
can often trace extended features to 0.5 mag fainter than this.}
\end{deluxetable}


\begin{deluxetable}{llrrrr}
\tablecaption{Measured Properties of the Arp 299 System}
\label{tab:global}
\tablehead{
\colhead{Quantity} &
\colhead{Units} &
\colhead{Total} &
\colhead{Disk\tablenotemark{a}} &
\colhead{Tail} &
\colhead{Companion\tablenotemark{b}} 
}
\startdata
Optical: \hfil \nl
~~~$L_B$\tablenotemark{c} &($L_{\odot,B}$)& $4.5\times 10^{10}$ 
& $4.3\times 10^{10}$ & $1.6\times 10^9$ & $2\times10^9$ \nl
~~~$L_R$\tablenotemark{c}  &($L_{\odot,R}$)& $4.9\times 10^{10}$
& $4.7\times 10^{10}$ & $1.2\times 10^9$ & $2\times10^9$ \nl
~~~$B$	& (mag)   & 12.31 & & & 18 \nl
~~~$R$	& (mag)   & 11.04 & & & 17 \nl\nl 
{\hi}, measured: \hfil \nl
~~~Velocity Range\tablenotemark{d} & (km s$^{-1}$) & 2800---3380
& 2910---3350 & 3020---3210 & 3230---3250 \nl
~~~$\int S dv$ & (Jy km s$^{-1}$) & 17.7 & 11.7 & 6.1 & 0.14 \nl
~~~\MHI\tablenotemark{e}& ($M_\odot$)  & $9.6\times10^9$ & $6.3\times10^9$ 
& $3.3\times10^9$ & $8\times10^7$ \nl
~~~\MhLb\    & (\ML) & 0.2 & 0.15 & 1.6 & 0.04 \nl
~~~\MhLr\    & (\ML) & 0.2 & 0.13 & 1.7 & 0.04 \nl\nl
{\hi}, corrected\tablenotemark{f}: \hfil \nl
~~~$\int S dv$ & (Jy \kms) & 21.7 & 15.7 \nl
~~~\MHI\   & ($M_\odot$)  & $1.18\times10^{10}$ & $8.5\times10^9$ \nl
~~~\MhLb\    & (\ML) & 0.3 & 0.20 \nl
~~~\MhLr\    & (\ML) & 0.3 & 0.18 \nl
\enddata
\tablenotetext{a}{Disk values include contribution from minor axis clump 
and western plume}
\tablenotetext{b}{HI detected companion located at 
$\alpha_{1950}= 11^h 24^m 28\fs0, \, \delta_{1950}= +58^\circ 48^\prime 
05^{\prime\prime}$ and $v_{\rm hel}=$~3245~\kms.}
\tablenotetext{c}{Luminosities calculated assuming $M_{\odot,B}$=+5.48 and 
$M_{\odot,R}$=+4.31. Values for disk are measured to \mB=26\msqas, 
\mR=25.5\msqas. Values for tail are measured after removing point
sources. The values for the companion are estimated from the DSS 
image and are very approximate; see text.}
\tablenotetext{d}{Range of {\hi} velocities (heliocentric). Taken from the
first-moment image except for the total, which is the range seen
in absorption.  The uncertainty is $\pm$5 \kms.} 
\tablenotetext{e}{Integrated {\hi} mass, calculated using
\MHI\ = $2.356\times10^5 M_\odot\,\Delta^2\,\int S dv$, where $\Delta$ is
the distance in Mpc, and $\int S dv$ is the integrated {\hi} emissivity, 
in Jy \kms. Following Sanders, Scoville \& Soifer (1991), we adopt 
48 Mpc as the distance to Arp 299.}
\tablenotetext{f}{Absorption corrected value; see text.}
\end{deluxetable}


\begin{deluxetable}{llrrr}
\tablecaption{Global Quantities}
\label{tab:progen}
\tablehead{
\colhead{Quantity} &
\colhead{Units} &
\colhead{Total}&
\colhead{IC 694} &
\colhead{NGC 3690} 
}
\startdata
$L_B$ &  $L_{\odot,B}$ & $4.5 \times 10^{10}$ & $2.8\times 10^{10}$ & $1.7\times 10^{10}$ \nl
$M_{HI}$  & $M_\odot$ & $1.18\times 10^{10}$ &  $6\times 10^9$ &  $6\times 10^9$ \nl
$M_{H_2}$\tablenotemark{a} & $M_\odot$ & $1.7\times 10^{10}$
& $1.1\times 10^{10}$ & $6\times 10^9$ \nl
$M_{gas}$\tablenotemark{b} & $M_\odot$ & $3.3\times 10^{10}$ &
$1.9\times 10^{10}$ &
$1.4\times 10^{10}$ \nl
$M_{HI}/L_B$ & $M_\odot \, L_{\odot,B}^{-1}$ & 0.3 & 0.2 & 0.4  \nl
$M_{H_2}/L_B$ & $M_\odot \, L_{\odot,B}^{-1}$ & 0.4 & 0.4 &  0.4 \nl
$M_{gas}/L_B$ & $M_\odot \, L_{\odot,B}^{-1}$ & 0.7 & 0.7 &  0.8  \nl
$M_{H_2}/M_{HI}$ & & 1.4 & 1.8  & 1.0 \nl
$L_{IR}$\tablenotemark{c} & ($L_{\odot,bol}$)& $7.9\times 10^{11}$ \nl
$L_{IR}/L_B$ &       & 18 \nl
$L_{IR}/M_{H_2}$ & ($L_{\odot} \, M_\odot^{-1}$)  & 69 \nl
\enddata
\tablenotetext{a}{From Casoli et al.~(1999), converting to our
distance and a CO-to-$H_2$ conversion factor of 
$3\times10^{20}\,{\rm cm^{-2}\,(K\, km\, s^{-1})^{-1}}$. 
We have equally divided the gas located at the overlap region 
$C-C^\prime$ between IC 694 and NGC 3690, and
assigned the ``widespread'' emission to IC 694.}
\tablenotetext{b}{Total gas mass corrected for the presence of {\sc H}e:
$M_{gas}=M_{H_2} + 1.34\times M_{HI}$}
\tablenotetext{c}{Far infrared luminosity over the range
8---1000$\mu$m, from Sanders et al.~(1991).}
\end{deluxetable}

\begin{table}
\tablenum{1}
\dummytable\label{tab:HIobs}
\end{table}

\begin{table}
\tablenum{2}
\dummytable\label{tab:OPTobs}
\end{table}
 
\begin{table}
\tablenum{3}
\dummytable\label{tab:global}
\end{table}

\begin{table}
\tablenum{4}
\dummytable\label{tab:progen}
\end{table}

\end{document}